\documentclass[conference]{IEEEtran}
\usepackage{cuted}
\usepackage{graphicx}
\begin{document}
\title{Ultra High-Energy Interaction of CR protons }

\author{\authorblockN{Tadeusz Wibig}
\authorblockA{Univ. of {\L }\'{o}d\'{z} and
So\l tan Inst. Nucl. Stud. \\
Uniwersytecka 5, 90-950 {\L }\'{o}d\'{z}, Poland., \\
Email: wibig@zpk.u.lodz.pl}
 }
\maketitle

\begin{abstract}
The recent Auger results suggest that although coincidences of arrival directions with 'nearby' AGN and HiRes discovery of the GZK cut-off indicate protons, the measured longitudinal propagation characteristics indicate heavy nuclei, if the conventional interaction model is correct. Something has to change! Our own view is that it is possible that the AGN -implied proton identification is not correct and that the extra galactic particles are, in fact, mainly 'heavies', in which case the interaction problem goes away. However, here we assume that the particles ARE protons and examine the possible consequences. Parameters discussed include the interaction mean-free-path, inelasticities and 'exotic' possibilities.
\end{abstract}

\section{Introduction}
Even a very brief review of connections between the Cosmic Ray (CR) physics and physics of high energy interactions has to start from the origin of the High Energy Physics (HEP) which has taken place high in the atmosphere where the multiparticle production processes involving  CR particles have been seen in nuclear emulsions. 

The second important conjunction point is associated with the first and simplest statistical model describing statistically these phenomena proposed by Fermi which was falsified also in the emulsion balloon experiments showing the asymmetry known today as the jet structure. 

Then, for some time, it was believed that the pre-ISR (Intersecting Storage Rings) HEP has reached an asymptotia and nothing unexpected appear when the energy rise. The cross sections were looking stable, according to the theory, as well as average transverse momenta, which were close to the Hagedorn limit of the highest ever possible temperature \cite{hage}. However, even in this pacific
time of the end of '60 there were suppositions, experimental indications
(see, e.g., \cite{sigmycr}) from cosmic ray physics where the energies were orders of magnitude higher than available for accelerator HEP that the Extensive Air Showers (EAS) do not look like they should, if nothing unexpected appears there.   
    This suspicions were undoubtedly confirmed by ISR. The cross sections started to increase, the transverse momenta distributions rise tails showing -- just unexpected. The ISR revolution initiated, to some extent, by the CR discoveries, drove to a fall of the thermodynamic, Hagedorn interaction picture (not entirely of course) and an ascent of the parton/quark hadronization related to the name of Feynman \cite{feynman}. 

    The Feynman scaling worked well and the quark hadronization picture, beautiful and simple satisfying everyone -- almost everyone. The debate concerning the Feynman scaling (in the forward region) initiated again in the cosmic ray community was resolved (with the certainty possible in colliding experiment when the very forward region is unseen) by the next machine, the Super Proton Synchrotron (SPS). This happened about the a quarter of century ago, just around the time when there was the European Cosmic Ray Symposium in Kosice for the first time.

    These four facts are of course a subjective choice, but, I think, that they ought to find considerable places in any list of the interaction points between the CR and HE Physics.

    The first remark I would like to underline in this introduction is: the long time has been passed since last event on the list. 

    The second point is related to attempt to answer the question, why this is so. Is it true that CR physics really has no discovery potential for contemporary HEP? The wishful answer is, of course, ``not!''

   In this paper we would like not only to diagnose the situation and propose the treatment, but also to show an example of the effect of such therapy. We will show the prediction to be tested in the future (not so very far future, indeed) related to discoveries recently announced by the Pierre Auger Observatory \cite{paorec} and HiRes \cite{hiresxmax}.

\section {CORSIKA}
The last decade of the previous century, was very successful in the particle physics, mainly in the theory. In the field of cosmic rays the attempt was made to answer one of the most important questions: about the nature of 'the knee' in the CR energy spectrum. 

The question was stated precisely in Kernforschungszentrum Karlsruhe and to answer it the construction of the new, powerful array which could be able to simultaneously measure as many shower characteristics as possible seems to be necessary . All particle components: electromagnetic, penetrating and hadronic carry the information needed to find the CR composition on the event-by-event basis.
    The array KASCADE (KArlsruhe Shower and Core Array DEtector) started to produce physical results in late '�90. Apart from experimental results there was another great achievement of the Karlsruhe group 
-- the shower simulation program CORSIKA (COsmic Ray SImulations for KAscade) \cite{corsika}.
The importance of the CORSIKA code was not only in its detailedness, exactness and completeness, but mostly in that that it was made widely available, user-friendly and clearly documented. The helpful assistance, extensive debugging with continuous significant improvements of physics involved, made the CORSIKA a kind of a standard tool to be used by different experimental groups and theoreticians to explore and compare results on cosmic rays also in the energy regions far from initially imposed, 'the knee' region around $10^{15}$--$10^{16}$ eV. These energies are not only of special interest for studies of the acceleration/confinement of CR in the Galaxy, but it also coincide, almost, with the highest accelerator energies available at that time. The high-energy interaction models which are essential for the EAS development were tested and adjusted to the SPS and Tevatron data.
 
Originally CORSIKA consists of only one high-energy interaction model, the Dual Parton picture inspired model of Capdevielle, HDPM \cite{jnc}. Then other models available on the marked were build-in to the CORSIKA structure as an options to be chosen by the user, simply by switching an appropriate flag in the steering cards.
   The history of major improvements is listed below.

\vspace{.2cm}
\begin{tabular}{ll}
1989 & CORSIKA 1.0 \\
& \ \ \ SH2C-60-K-OSL-E-SPEC (Grieder, 1980)
\\
\multicolumn{2}{r}{main structure, isobar model}\\
& \ \ \ ESKAR (HDPM) (Capdevielle, 1987)\\
\multicolumn{2}{r}{high-energy interaction}\\
& \ \ \ EGS4 (Nelson {\it et al.} 1985)
\\
\multicolumn{2}{r}{electro-gamma shower}
\\
& \ \ \ NKG (Capdevielle, 1989)
\\
\multicolumn{2}{r}{analytic EM subshowers}\\
1994 &CORSIKA4.006\\
& \ \ \ GHEISHA (Feselfeld, 1985)\\
& \ \ \ VENUS (Werner, 1993)\\
1997 &CORSIKA 5.20
\\
& \ \ \ SIBYLL (Fletcher, Geisser {\it et al.} 1994)\\
& \ \ \ QGSJET (Kalmykov {\it et al.} 1993)\\
& \ \ \ DPMJET (Ranft, 1995)\\
2000&CORSIKA 6.00
\\
& \ \ \ NEXUS (Drescher {\it et al.} 2001)\\
& \ \ \ UrQMD (Bleicher {\it et al.} 1999)\\
2004 &CORSIKA 6.20\\
& \ \ \ FLUKA (Fasso, Ferrari {\it et al.} 2001)\\
2006 &CORSIKA 6.535\\
& \ \ \ EPOS (Werner, Liu and Pierog 2006)\\
\end{tabular}

\noindent
(detailed references can be found in Ref.\cite{corsika2})

The most actual (June 27, 2008) version CORSIKA  has a number 
6.735 and consists of:

\vspace{.2cm}
\begin{tabular}{ll}

EPOS -- 161   &Fluka 2006
\\
DPMJET - II.55& GHEISHA\_2003d
\\
NEXUS -- 397&URQMD -- 1.3cors
\\
QGSJET01C&HERWIG 6.510
\\
QGSJET -- II -- 3~~~~~~~~~~
\\
SIBYLL
\\
VENUS 4.12/5
\end{tabular}

The overlapping the time periods of the weakening of the CR and HEP relations and the proliferation of the CORSIKA, could be a pure coincidence, but it do not have to. In my, very personal opinion,
the development of the CORSIKA, paradoxically, slowdown, and even, to some extent, suspend further CR driven development of high-energy interaction modeling. The multiparticle production codes introduced in CORSIKA overwhelmed all the others. Consecutive progress (see, e.g.,\cite{corsika22}) in these particular codes made by the original Authors is important, but the last years show that the solution is still far-off. ‘The solution’ means the correct description of EAS with the lack of contradictions from the accelerator measured characteristics. For recent example see, e.g., the paper by the KSCADE Collaboration Ref.~\cite{apel}.

\section {The simple analytic solution}
In this paper we would like to study EAS initiated by the particles (protons) of energies of $10^{18}$ eV and above. Such events are rare, thus the surface detectors to study them has to be spread over large area (if one wants to use the surface detector, there are other techniques available). The situation is qualitatively different then the one around 'the knee' where the KASCADE measured each shower in details for its hadron, muon, and electromagnetic constituents. Such multicomponent and accurate measurements could give a profitable information about the interaction properties, but, on the other hand, this information is sometimes hard to handle, as it was just mentioned. For very high energies only some characteristics of the shower can be measured with the enough accuracy. The most important is the shower longitudinal development, (given experimentally as, e.g., the distribution along the shower axis of the fluorescent light emitted when the charged shower particles excite air molecules). This distribution could be measured to some extent only. The normalization (total number of particles) which is related closely to the primary particle energy, and first moment (given usually as $x_{\rm max}$ -- the position of the maximum number of the particles in the cascade) are the parameters available for further study. The recent measurements of PAO \cite{paoxmax} and HiRes \cite{hiresxmax} together with the older from Fly's Eye \cite{fexmax} and Yakutsk \cite{yakutskxmax} gives the whole information of the average value of $x_{\rm max}$ as a function of estimated primary particle energy one can get. 

    The interpretation of shower data in 'the �knee' region needs simulation programs as complicated as CORSIKA to utilize the gathered information fully. To explore only the average $x_{\rm max}$ data we can use a much simpler code.

   If we denote by $N_{\{p,\pi,K,\mu\}}(E,\:x)$ number of particles of the type $p,\pi,K $ or $\mu$ crossing the depth $x$ with energies between $E$ and 
$E+dE$, then $N$ should evolve traversing the slab of matter of the depth $dx$ 
\begin{itemize}
\item[-]{decreasing according to some probability of interact or decay within 
$dx$}, 
\item[-]{and 
possibly increasing by the average amount of particles produced by the
higher energy particles entering the slab.}
\end{itemize}
The only parameters of the respective system of integral-differential transport equations are probability of interaction (decay) and the inclusive energetic characteristics of multiparticle production processes.
The decay constants and branching ratios needed for our purposes are known very well so we'll concentrate on interactions hereafter.

The interaction probabilities are given by the cross sections (we'll also leave aside the problem of geometry and atmosphere modelling). They are of course not measured for the energies of our present interests, so we must extrapolate the low energy values. Changing the cross sections we can obviously control the rate of the shower development and move a shower maximum where it we wish it to be. But there are some constrains, and in recent years they become quite strong. We will discussed this point later.

The inclusive energy distributions are known up to SPS (and Tevatron) energies and the knowledge is not as good as we wish it to be. Lower energy experiments (around $\sqrt{2}\approx 20 $ GeV) with stationary target produced more precise data and, what is more important, they cover the very forward region, which in fact controls a development of a cascade in thick media.
The information we get is limited and we need to follow some more or less elaborate models to extrapolate it.

\subsection{HDPM}

\begin{figure}[th]
\centerline{
\includegraphics[width=7.7cm]{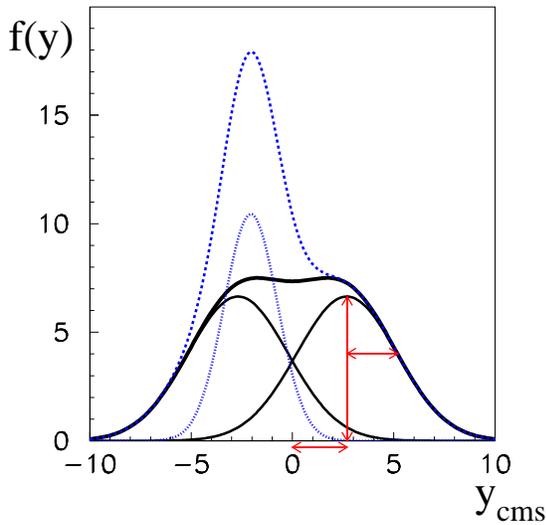}}
\caption{Definition of three main parameters of HDPM. The dashed (blue)
components shows the contribution introduced by the air target nucleus
additional chains. 
\label{hdpmpars}}
\end{figure}

\begin{figure}[th]
\centerline{
\includegraphics[width=7.7cm]{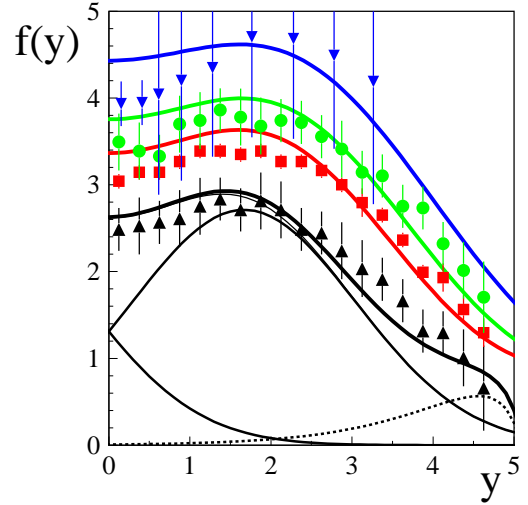}}
\caption{The SPS and Tevatron data on (pseudo)rapidity distribution
compared with HDPM prediction with standard values of the model parameters
\label{hdpmsps}}
\end{figure}

The high-energy interaction model used in the CORSIKA program from the  very beginning was the so-called HDPM model \cite{jnc}. This phenomenological parametrization of the available data in the form expected by the two-chain structure expected in the Dual Parton jet hadronization pictures. In the proton-proton case it assumes on average that secondary particle rapidities are distributed according to two chains (jets) described by Gaussian (in rapidity). The widths, heights and positions of these Gaussians, as shown in the 
Fig.~\ref{hdpmpars},  are three most important parameters of the model. They are adjusted to the data up to SPS energies as shown in Fig.~\ref{hdpmsps} and extrapolated smoothly. The parameters describe roughly the inelasticity and the average multiplicity. As the energy increase both increase as it is shown in Figs.~3 and 4.

\subsection{Wdowczyk and Wolfendale scaling breaking model}

Another interaction model which we wont to examine in the present work is the scaling breaking model of Wdowczyk and Wolfendale (WW) (\cite{ww}). It has been proposed to described the CR data at the beginning of '70. Formally it is a generalization of the Feynman scaling idea given by:
\begin{equation}{
{2E \over \sqrt{s}}\: {d^2 \sigma \over dp_\| dp_\bot }~
=~
f \left( 
 {p_\| \over p_{\rm max} }
,~p_\bot \right)
}~~~,
\end{equation}
where $\sqrt{s}$ is the interaction c.m.s. energy $E$, $p_\|$ and $p\bot$ are energy, longitudinal and transverse momentum of created particle and $p_{\rm max}$ is the maximum momentum which can be taken by the particle of the particular type (mass).

The scaling, as it is widely known, was suggested in Ref.\cite{feynman}\footnote{It is interesting to note that the same idea appeared in the CR physics twenty years earlier in the paper by Heitler and J\'{a}nossy:
``{\bf 
$\mathbf{
\Phi(E',\:E)~=~\Phi \left({E' \over E }\right) {d \:E' \over E }
}
$
shall be a function of ${{E' \over E }}$ only.}''
(where E is the incoming particle energy, $E’$ energy of the secondary particle and $\Phi$ is the probability density of producing particle $E’$).}

The WW modification is in the additional term

\begin{equation}{
{2E \over \sqrt{s}}\: {d^2 \sigma \over dp_\| dp_\bot }~
=~
\left( {s \over s_0}\right)^\alpha ~
f \left( 
 {p_\| \over p_{\rm max} }~
\left( {s \over s_0 }\right)^\alpha 
,p_\bot \right)
}~~~.
\end{equation}
If the parameter of the model $\alpha$ is equal 0 then the Feynman scaling is restored. For $\alpha$ = 0.25 the interaction multiplicity increases as $E^{1/4}$ and, in the sense,  this value can be treated as the upper, thermodynamical limit. The value of $\alpha$ originally introduced to the WW model is 0.13. This value is based on the interpolation of the $x_F=p_\| /p_{\rm max}$ distributions between $sqrt(s) \approx 10$ GeV and ISR energies. The increase of the central rapidity density reported in Ref.\cite{alner} suggests $\alpha = 0.105$. There are evidences from the high energy cosmic ray data that alpha could be even as big as 0.18.

For the EAS description the WW model in its version of the mid '80 was improved by introducing 
partial inelasticities $k\left(s,\:s_0\right)$ the slowly changing functions, which can give a better description of the production of different kind of secondaries. The model predicts, for example, the increasing role of the production of the barions, and this was realized by the power-law correction factor $k$ with an index of 0.042.   

\begin{equation}{
{2\pi \over \sigma_{inel.}}\: {d^2 \sigma \over dp_\| dp_\bot }
={k\left(s,\:s_0\right) \over E}
\left( {s \over s_0}\right)^\alpha 
f \left( 
 {p_\| \over p_{\rm max} }
\left( {s \over s_0 }\right)^\alpha 
,p_\bot \right)
}
\end{equation}

The agreement of the WW model description of the SPS data was shown e.g. in Ref.~\cite{alner}. 

\section{Results of HDPM and WW models at high energies}
Results of the comparison of the standard HDPM an the WW model at very high energies is given below. For the value of $\alpha$ we used for the moment 0.105 as suggested in Ref.~\cite{alner}.

The change of the inclusive distributions, rapidity for HDPM and $x_F$ for the WW model, as the laboratory energy of projectile proton increases from $10^{14}$ to $10^{19}$ eV is presented in Fig.~3.

\begin{strip}
\centerline{
\includegraphics[width=7.7cm]{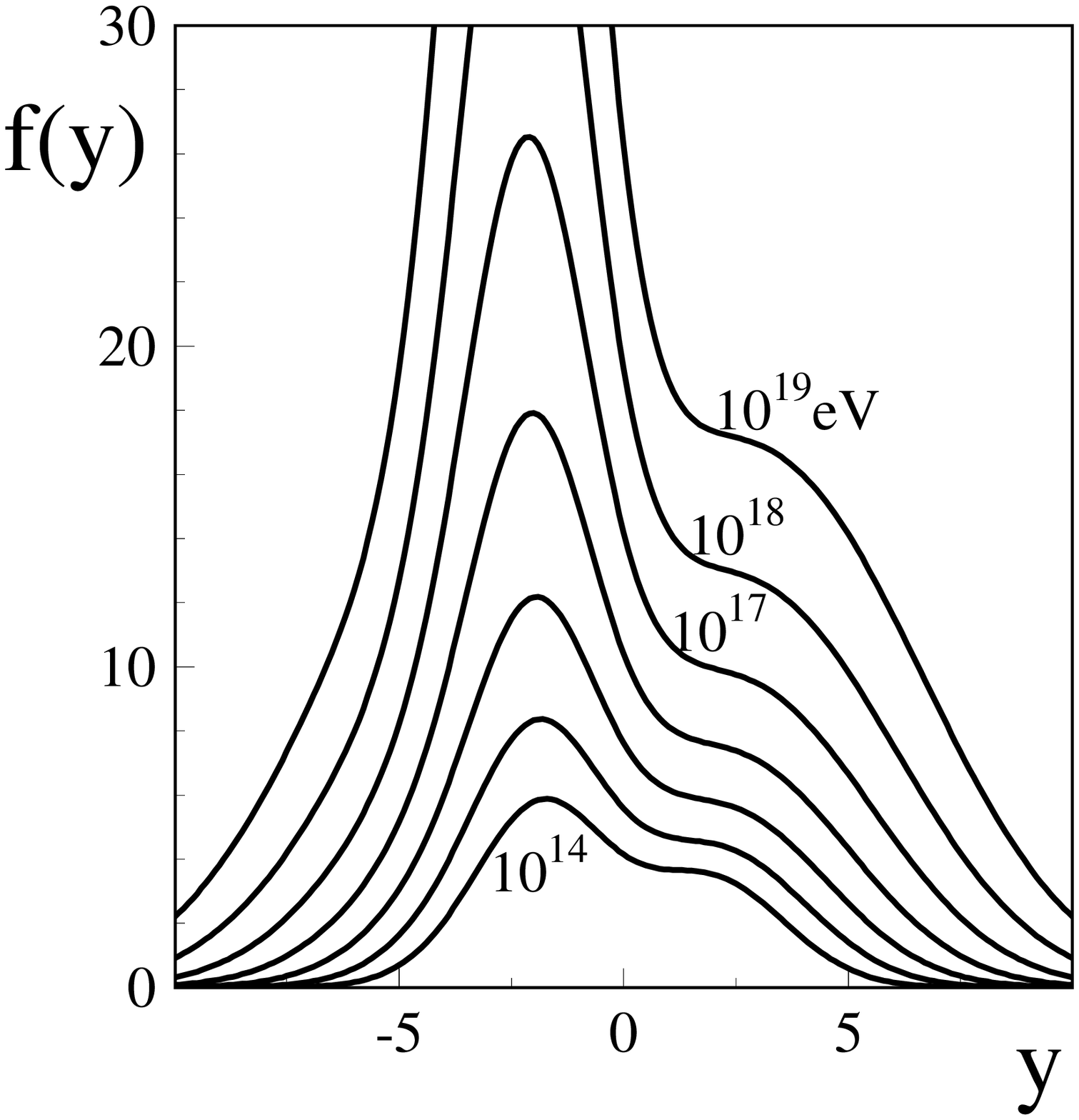}
\includegraphics[width=7.7cm]{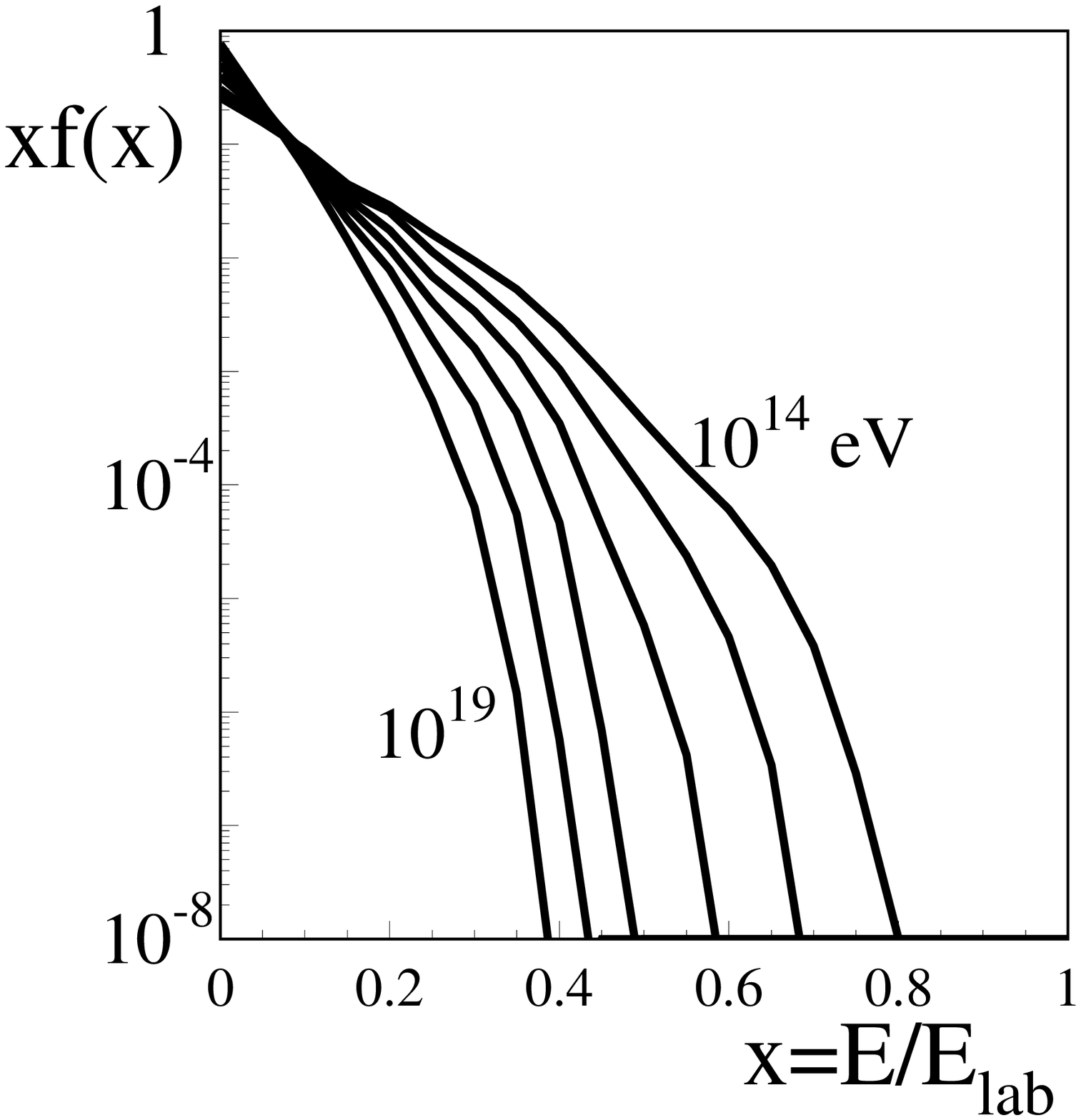}}
\footnotesize{Fig. 3. The change of the rapidity distributions of standard HDPM (left) and inclusive Feynman $x$ variable distribution of the WW model (right) with projectile proton laboratory energy.
}
\end{strip}

\setcounter{figure}{3}
\begin{figure}[th]
\centerline{
\includegraphics[width=4.5cm]{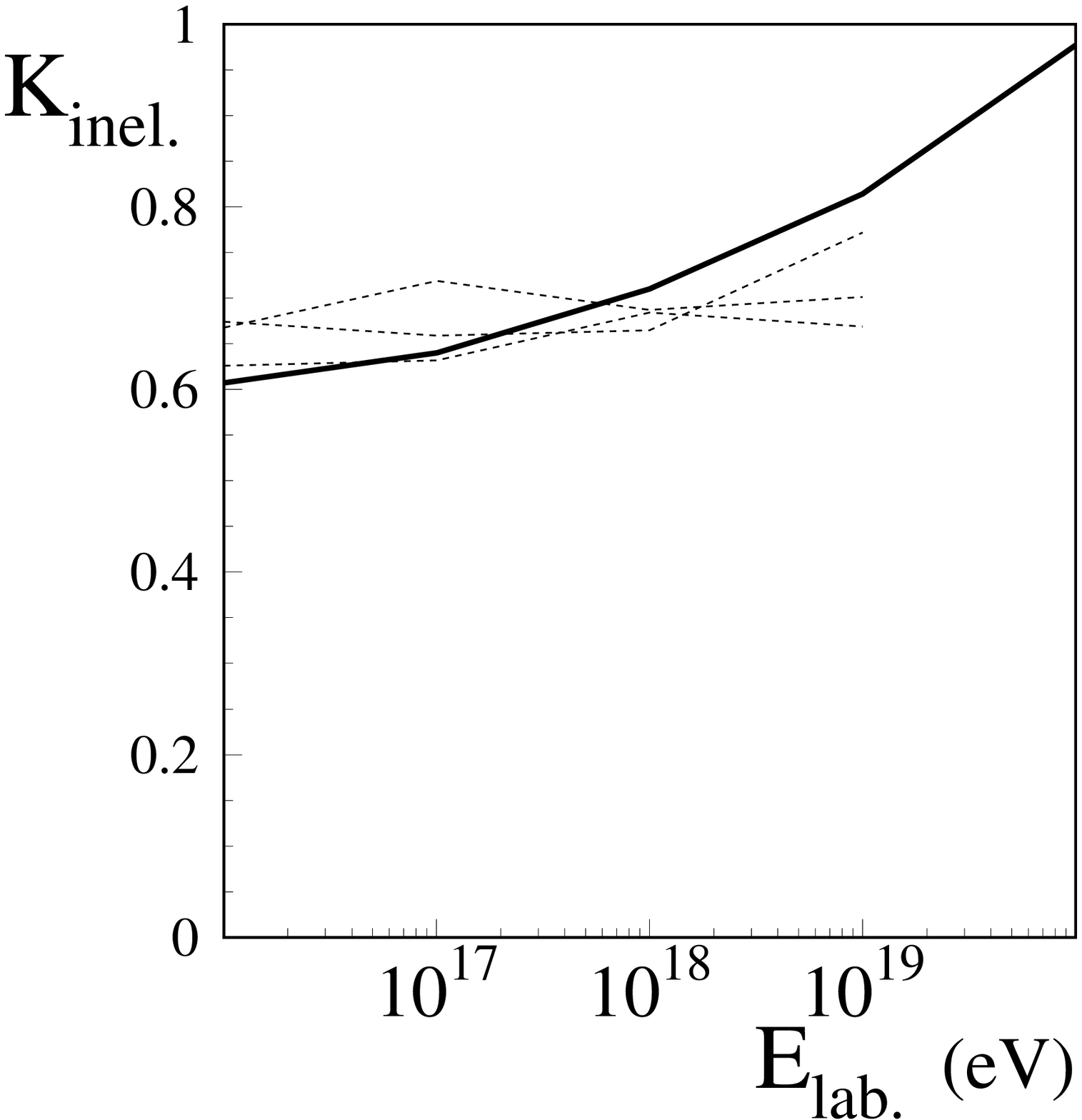}\hspace{-.5cm}
\includegraphics[width=4.5cm]{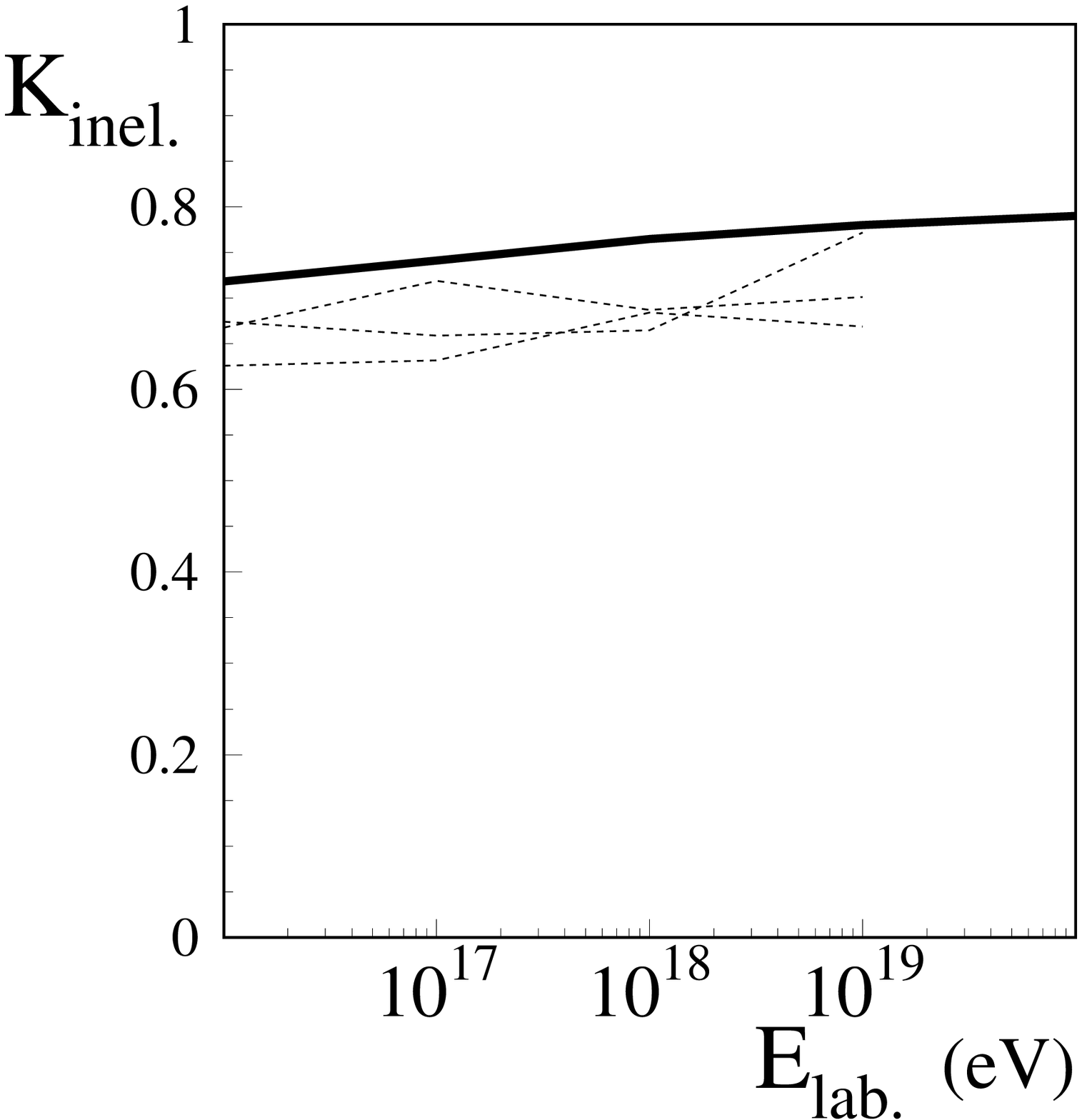}}
\caption{Average inelasticity in p-Air interaction as a function of the proton energy
for HDPM (left) and WW model (right). }
\end{figure}
The differences in interaction inelasticity and in average multiplicities are shown in Figs. 4 and 5. It is worth to noticed that they are not very big. It is hard to speculate on the basis of these figures about the  resulting difference in $x_{\rm max}$ position. Surprisingly it is substantial. It comes from the behaviour of the inclusive $x$ (or $y$ and correlation of it with $p_\bot$) distribution if the forward region rather than from global interaction characteristics. The WW model even with $\alpha $ = 0.105 brakes the Feynman scaling stronger than the HDPM change of the respective Gaussian widths.

With the help of fast analytic program the average position of maximum of the shower can be calculated very fast. Results are shown in Fig. 6 for both models. The thin dashed lines shows the results of other (SIBYLL, DPMJET, QGSJET) CORSIKA models (the same is shown by the dashed lines in Fig.~5).
These CORSIKA results in fact are well known from literature.

The lack of the direct data which could help to solve the problem about the real nature of the forward fragmentation. We can try in this paper to use the cosmic ray data and the assumption about the 'pure proton' composition above the ankle to look for a possible solution.

We will try to change an extrapolation of the model parameters logarithmically as, e.g., in Ref.~\cite{ulrich}.

\begin{equation}
f_{\rm new}(E)~=~f_{\rm old}(E) \left\{    
\begin{array}{lr}
 1   &       E \le 1 {\rm PeV} \\                                          
1 + \left( \delta_{{19}} - 1 \right)\:
{\lg (E) - 15\over (19-15)}  &E>1 {\rm PeV}
\end{array}
\right.~~~.
\label{corrfac}
\end{equation}

This mean that, up to the energies of SPS (roughly), where the models have been adjusted to the data, we do not change anything and for higher energies the approximations are corrected by a slowly varying factor determined by the value of the correction $\delta_{{19}}$ almost at the end of CR data, at $10^{19}$ eV.

This correction can be applied to multiparticle production model parameters as well as to the cross section values. But this has to be done with care.

\begin{strip}
\centerline{
\includegraphics[width=7.7cm]{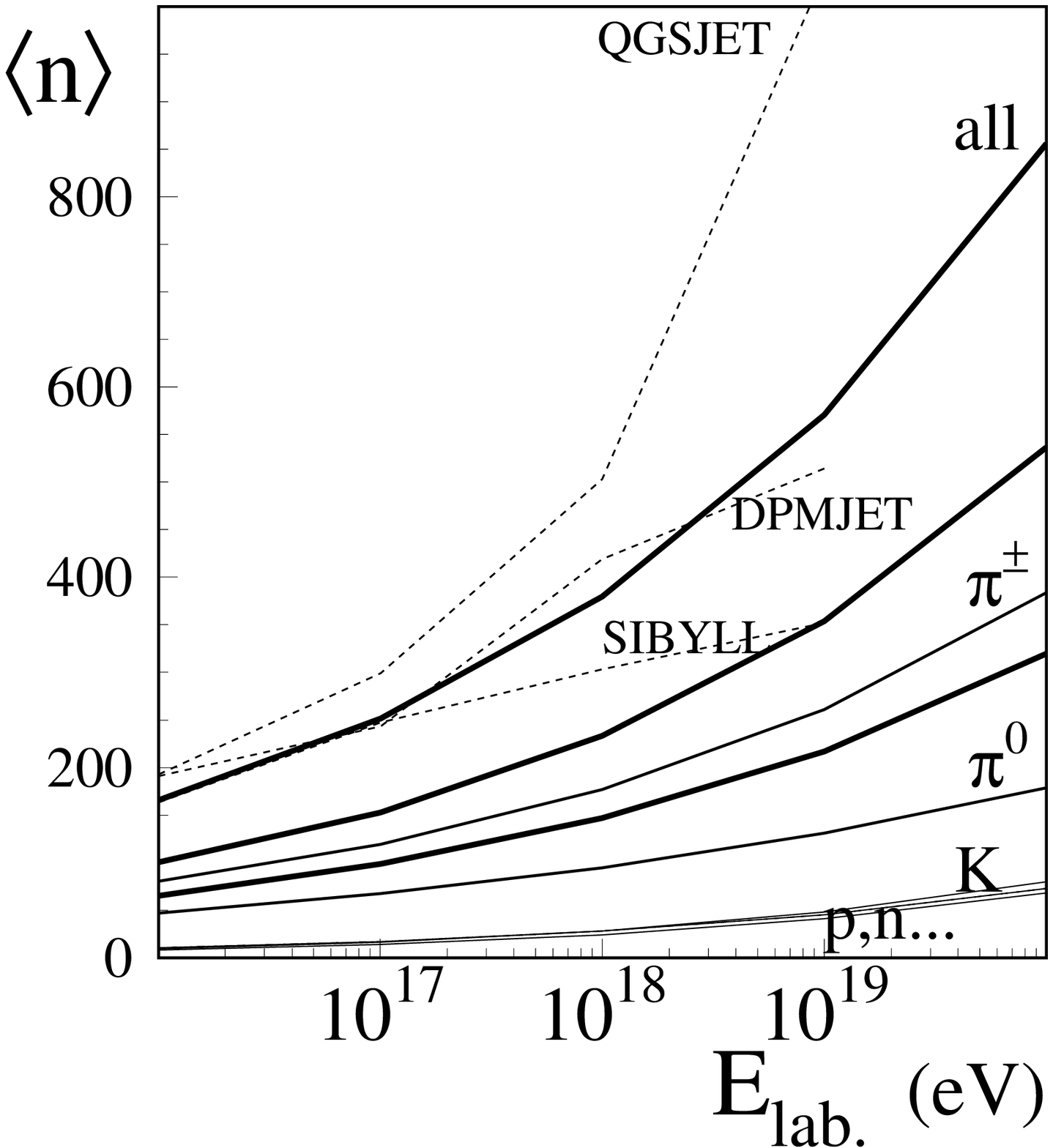}
\includegraphics[width=7.7cm]{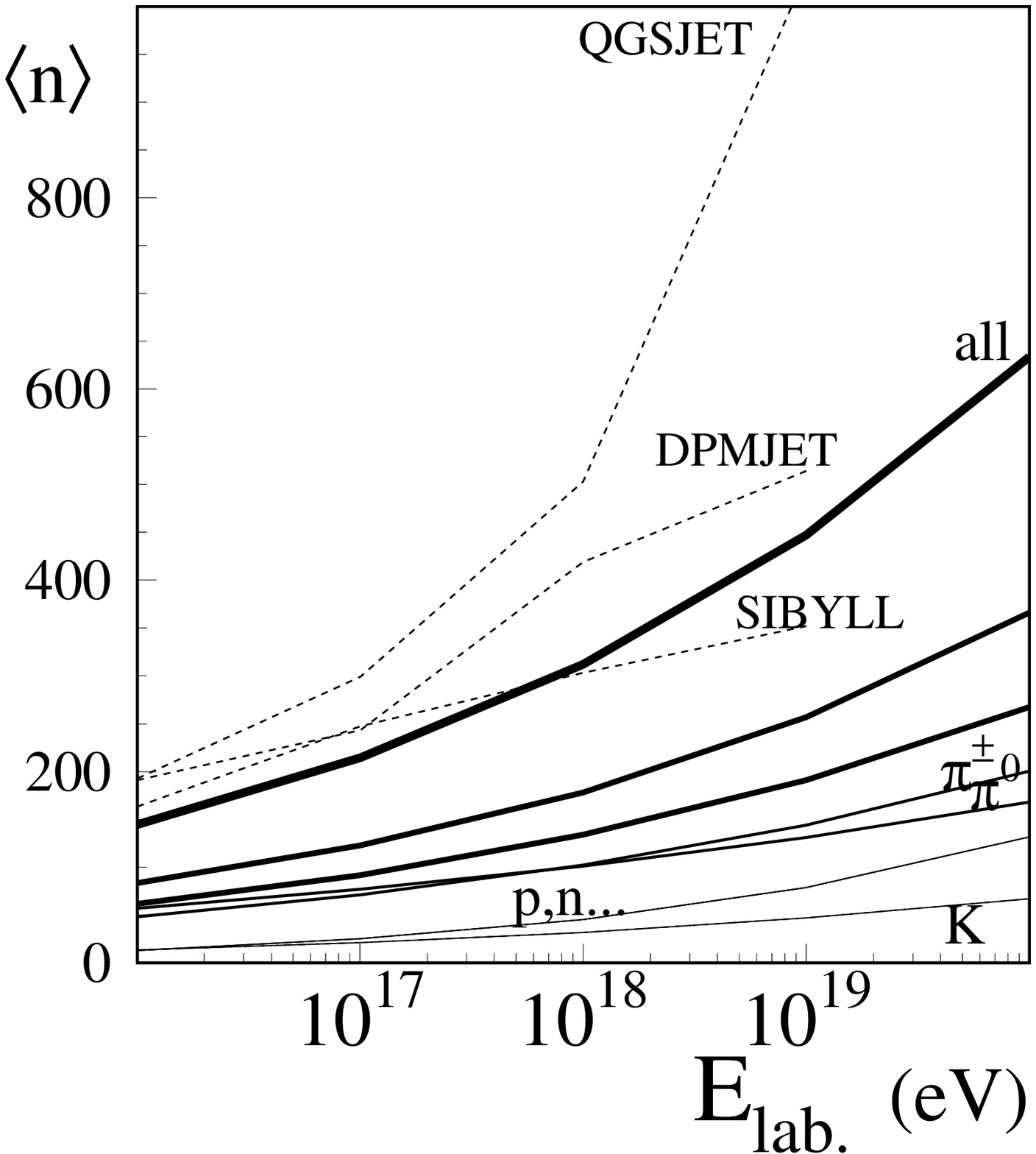}}
\footnotesize{Fig. 5. Average multiplicity in $p$-Air interaction as a function of the proton energy
for HDPM (left) and WW model (right). 
\label{hdpminel}}
\end{strip}

\begin{strip}
\centerline{
\includegraphics[width=7.7cm]{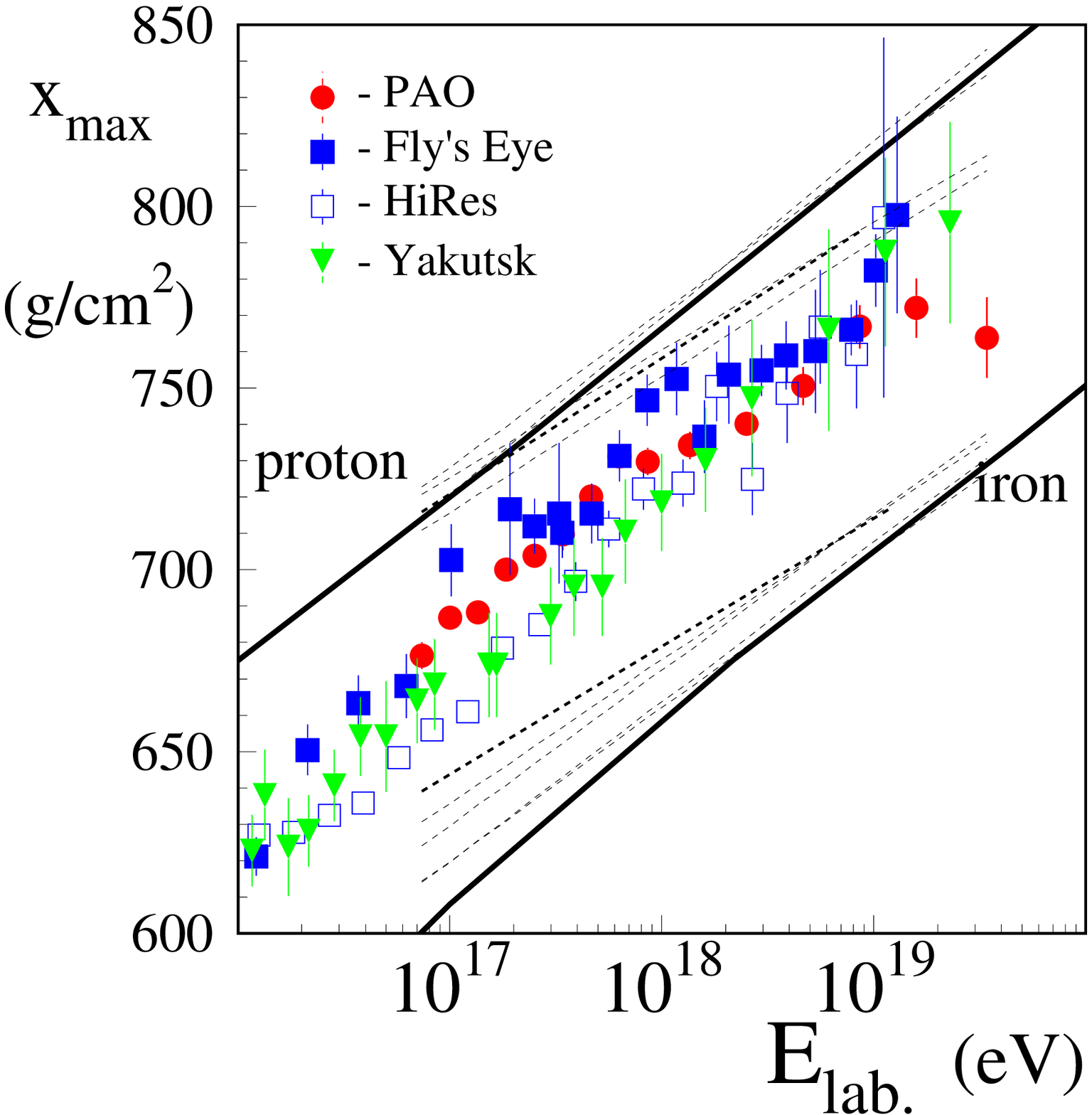}
\includegraphics[width=7.7cm]{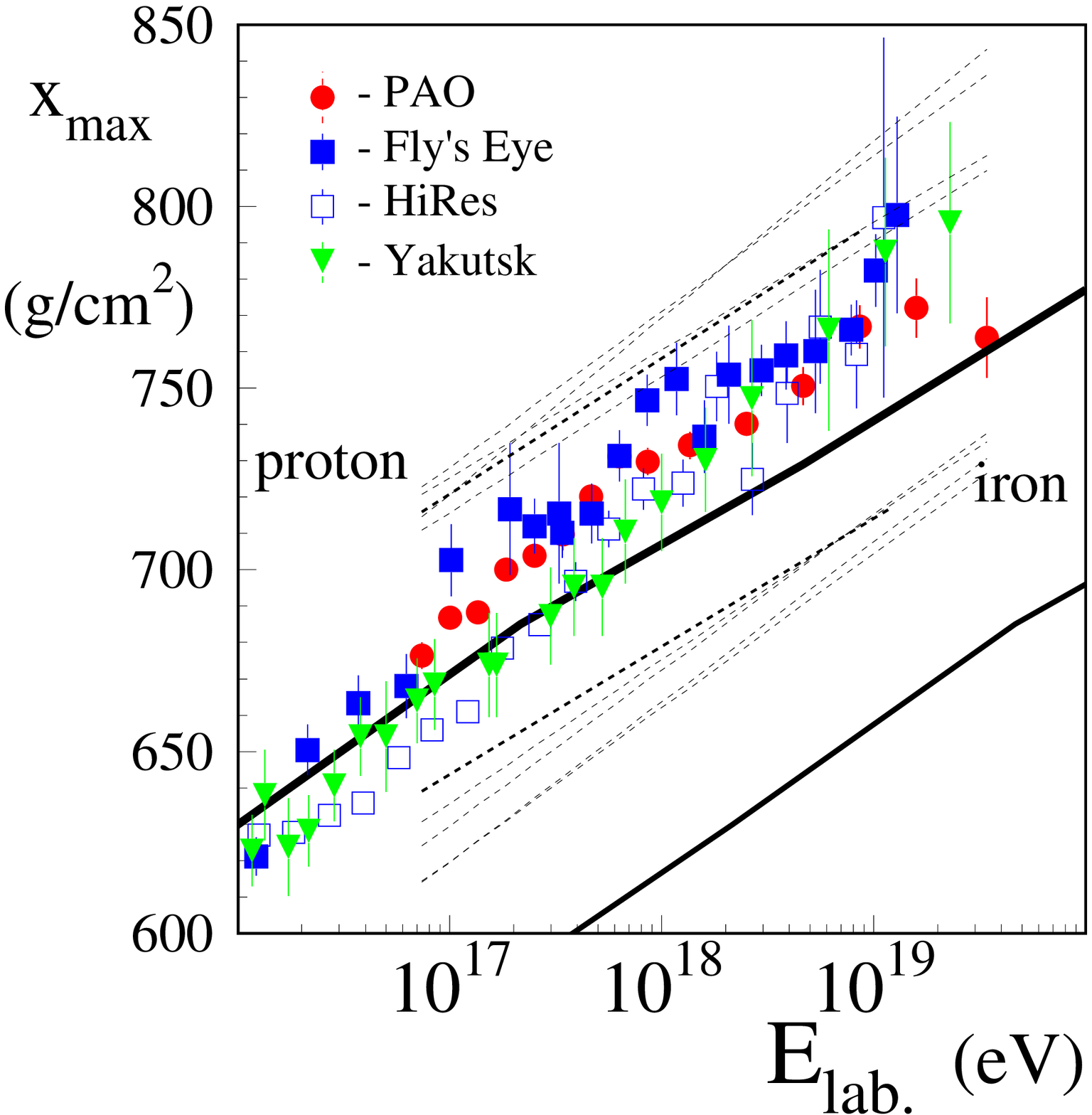}}
\footnotesize{Fig. 6. The prediction the proton induced shower $x_{\rm max}$ depths
calculated with HDPM (left and WW model (right) in a comparison with different measurement data.
\label{hdpmxmax}}
\end{strip}
\setcounter{figure}{6}

\section{Recent progress in cross section description}

The cross section involved in the EAS development is of course hadron-nucleus cross section. Before we discuss this complex case we would like to look closely at the interaction with single proton.

\subsection{Hadron-proton scattering cross section}

Recent years brought the significant progress in the theoretical description of the inelastic (elastic and total) cross section. It is based
on the optical picture
\begin{eqnarray}
{\sigma}_{\rm tot}~=~2\:\int\:\left[\:1\:-\:{\rm Re}\left (
{\rm e}^{i \chi ({b})}\right)\: \right]
d^2 {\bf b}~~,
\nonumber \\
{\sigma}_{\rm el}~=~\int\:\left| \:1\:-\:{\rm e}^{i \chi ({b})}\:\right| 
^2\:
d^2 {\bf b}~~,
\label{sigsig}
\nonumber \\
{\sigma}_{\rm inel}~=~\int\:1\:-\:\left| \:{\rm e}^{i \chi ({b})}\:\right| 
^2\:
d^2 {\bf b} ~~.
\end{eqnarray}
The phase shift $\chi$ is related to the scattering amplitude
by the two-dimensional Fourier transform
\begin{eqnarray}
1\:-\:{\rm e}^{i\chi({\bf b})}
~=~{{1} \over {2\: \pi \: i}}\int\:{\rm e}
^{-i{\bf b\:t}} S({\bf t}) d^2 {\bf t} ; \nonumber \\
S({t}) ~=~{i \over {2\: \pi \: }}\int\:{\rm e}
^{ i{\bf b\:t}} \left(
1\:-\:{\rm e}^{i\chi({\bf b})} \right)
d^2 {\bf b} .
\label{eq2}
\end{eqnarray}
Using the optical analogy one can interpret the $
1\:-\:{\rm e}^{i\chi({\bf b})}
$ function as a transmission coefficient
for a given impact parameter. Considering two colliding object we can
assume (for pure absorptive potential)
\begin{equation}
\chi({b})~=~i\: \omega(b)~=~i\: K_{ab}\int \: d^2{\bf b'}\:
\rho_a({\bf b})\rho_b({\bf b}\:+\:{\bf b'}) ,
\label{rho}
\end{equation}
\noindent
where $\rho_h$ is a particle's ``opaqueness''
(the matter density integrated
along the collision axis).

In the serial of papers by Block and coworkers \cite{sigblock}
the approximation of $\chi$ of the form inspired by QCD
\begin{eqnarray}
\chi(b,s)~=~ \xi_{qq}(s,b)+\xi_{qg}(s,b)+\xi_{gg}(s,b)~=~ ~~~~~~~~~~~~
\nonumber \\
 ~~=~
{\it i}\:\left[ \sigma_{qq} W(b; \mu_{qq})+
\sigma_{qg} W(b; \mu_{qg})+\sigma_{gg} W(b; \mu_{gg}) \right]
\end{eqnarray}
with
\begin{equation}{
W(b;\mu)~=~{\mu^2 \over 96 \pi} (\mu b)^3 \: K_3(\mu b)
}\end{equation}
give very good description of $pp$ and $p \bar p$ data. Assuming the vector meson dominance and the additive quark model, it could be used with the same parameters also for photon-proton and photo-photon scattering cross section calculations.


Another parametrization 
\begin{equation}
 \chi(s,b)  ~=~(\lambda(s)+i)\: \omega(b,s)
\label{lambdach}
\end{equation}
was proposed by P\'{e}rez-Peraza and collaborators in Ref.~\cite {perper}.

\begin{eqnarray}
\omega(b,s)=~C\:\left\{ 
E_1\:K_0(\alpha b) +
E_2\:K_0(\alpha b) +
E_3\:K_{\rm ei}(a b) +
 \right.
\nonumber \\
~~~\left.
E_4\:K_{\rm er}(a b) +~b\left[
E_5\:K_1(\alpha b) +
E_6\:K_1(\beta b) 
\right]\right\}
\end{eqnarray}
was fitted to differential elastic scattering data. The
consistency of ISR, SPS and Tevatron\footnote{Tevatron cross section is assumed to be  of 74.21$\pm$1.35, the weighted arithmetic mean of the E710 (72.8$\pm$3.1 mb), CDF (80.3$\pm$2.3 mb) and
a E811 (71.7$\pm$2.0 mb) values}
cross section within the framework of adopted model makes the eventual
confidence band narrow. 

In the paper by Ishida and Igi \cite{ishida}
the cross section of $K^\pm p$, $\pi^\pm p$, $p \bar p$ and $pp$
the scattering amplitude was parametrised in the form
\begin{equation}{
f(s)={s \over m^2}\left[ c_0+c_1 \log \left({s \over m}\right)+c_2 \log^2 \left({s \over m}\right) \right]
+ {\beta_{P'} \over m}
\left({s \over m}\right)^{\alpha_P'}
}\end{equation}
which, for high energies, leads to the saturation of the Froissart bound
\begin{equation}{
\sigma~ \simeq ~B\log^2 (s/s_0) ~=~
 \left(4 \pi \over m^2 \right) c_2\log^2 (s/s_0)
}\end{equation}
The universality of the value of $B$ found in Ref.~\cite{ishida} 
0.289$\pm$0.023, 0.351$\pm$0.036, 0.37$\pm$0.21 for 
$pp$ ($p\bar p$), $\pi^\pm$, and $K^\pm$, respectively, gives additional 
evidence that proposed picture is correct and extrapolations are 
thus strongly justified.

The very similar and self-consistent description of data on charged pion and proton (antiproton) projectiles 
has been given by Block and Halzen in Ref~\cite{blockhalz}. 
\begin{equation}{
\sigma(s)= c_0+c_1 \log \left({s \over m}\right)+c_2 \log^2 \left({s \over m}\right) 
+
{\beta_{P'}}
\left({s\over m}\right)^{\alpha'}~~~.
}\end{equation}
With some new tool of calculating 'the best fit' the 
necessity of $\log^2$ component is confirmed.

The $\log^2$ character of the cross section rise was argued for by COMPETE Collaboration in Ref.\cite{cudell}.

We would like to remind here briefly the result of Ref.~\cite{wibso} from 1998.

The cross sections for $K^\pm p$, $\pi^\pm p$, $p \bar p$ and $pp$
interactions were parametrized assumed geometrical scaling

\begin{figure}[th]
\centerline{
\includegraphics[width=7.7cm]{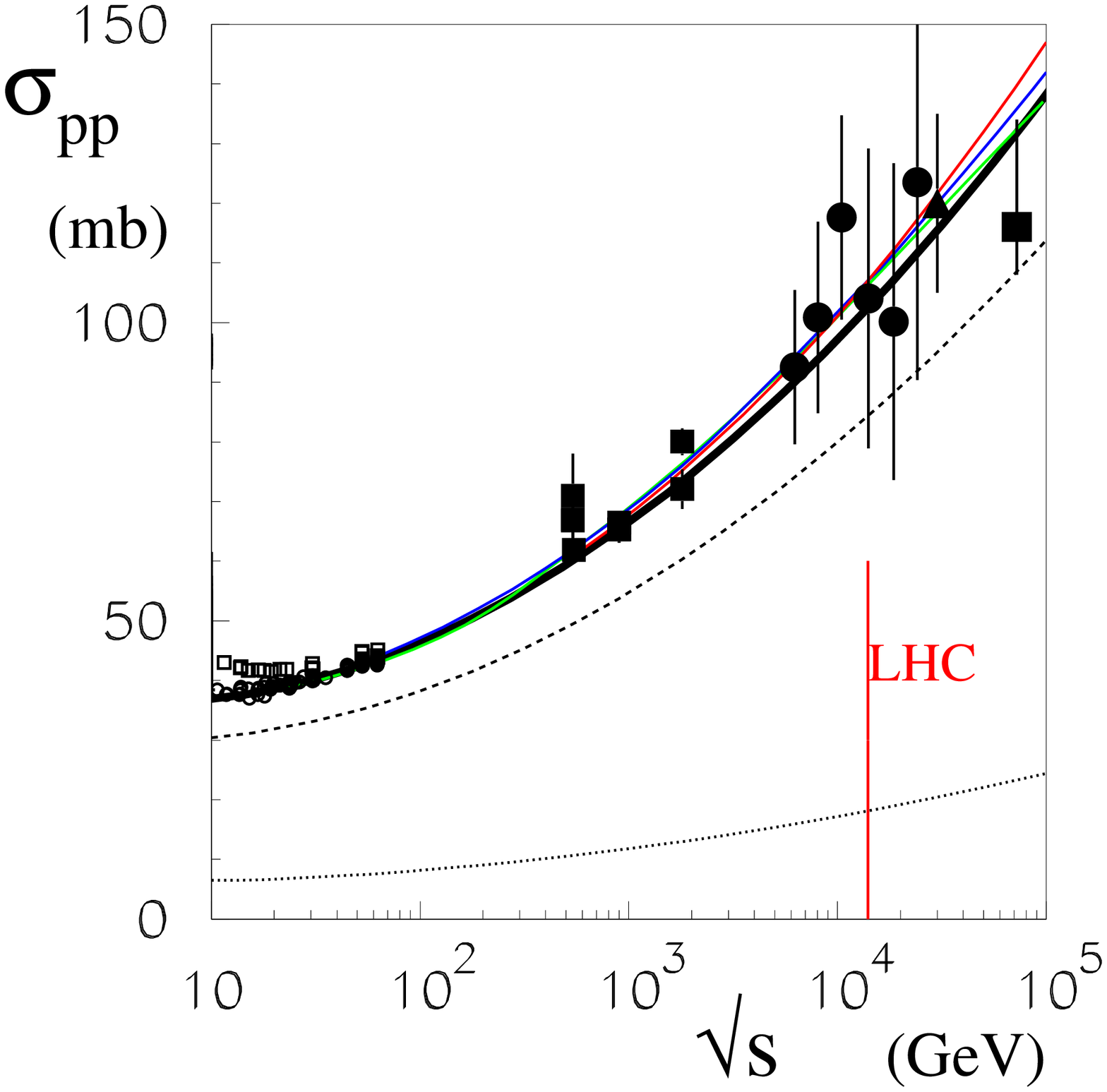}}
\caption{Proton-proton total cross section calculated using parametrisation Eq.(\ref{inel}) compared with the data from accelerator measurements an recalculated cosmic ray attenuation data points. Dashed and dotted lines represent the inelastic and elastic contributions. Dotted and dash-dotted (coloured) lines are the
results from other well known approximations \cite{blca,durpi,honda}.
\label{sigpp}}
\end{figure}

\begin{equation}
\omega(b,s) ~=~
\omega(\widetilde b ) ~~~~~{\rm with} ~~~~
\widetilde b
~=~  b \:  \left[{{\sigma_{\rm inel}(s_0)}
\over {\sigma_{\rm inel}(s)}} \right]
^{ \frac 1 2 } ,
\label{gsomega}
\end{equation}
where omega was calculated as a convolution of colliding hadron matter distributions
described with the help of only one parameter ($m_\pi$,$m_K$, $m_p$) for each interacting particle type.
\begin{equation}
\rho_h({\bf b})~=~\int d z {{ m_h} \over {8 \pi}} {\rm e}^{-m_h {\bf r}}
\label{pdens}
\end{equation}

Values of $\lambda$ and $\sigma_{\rm inel}$ are found:
\begin{equation}
\lambda(s)~=~{{0.077\: \ln (s/s_0)} \over {
1~+~0.18\: \ln (s/s_0)~+~.015\: \ln ^2 (s/s_0)}}~,
\label{lamb}
\end{equation}
\begin{equation}
{\sigma_{\rm inel}(s)} = 32.4~-~1.2 \ln (s)~+~0.21 \ln ^2 (s)~,
\label{inel}
\end{equation}
($s_0 = 500$ GeV$^2$). 
The resulting cross sections are given in Fig.\ref{sigpp} together with 
other approximations by Block and Cahn \cite{blca}, Durand and Pi \cite{durpi} and Honda 
parametrization using Akeno data \cite{honda}. 

\subsection{Proton-nucleus and nucleus-nucleus interaction cross section}
The scattering of particle on the close many-particle system (nucleus)
can be treated as a superposition of individual interactions each with a specific 
phase shift.
The overall phase shift for incoming wave is a sum of all the two-particle
phase shifts.
\begin{equation}
\chi_A(b,\: \{{\bf d}\})~=~\sum_{j=1}^A \: \chi _j ({\bf b}\: -\: {\bf d}_j)
\label{chiskla}
\end{equation}
Eq. (\ref{chiskla}) is the essence of the Ref.\cite{glauber}
and defines the Glauber approximation.

\begin{figure}[th]
\centerline{
\includegraphics[width=7.7cm]{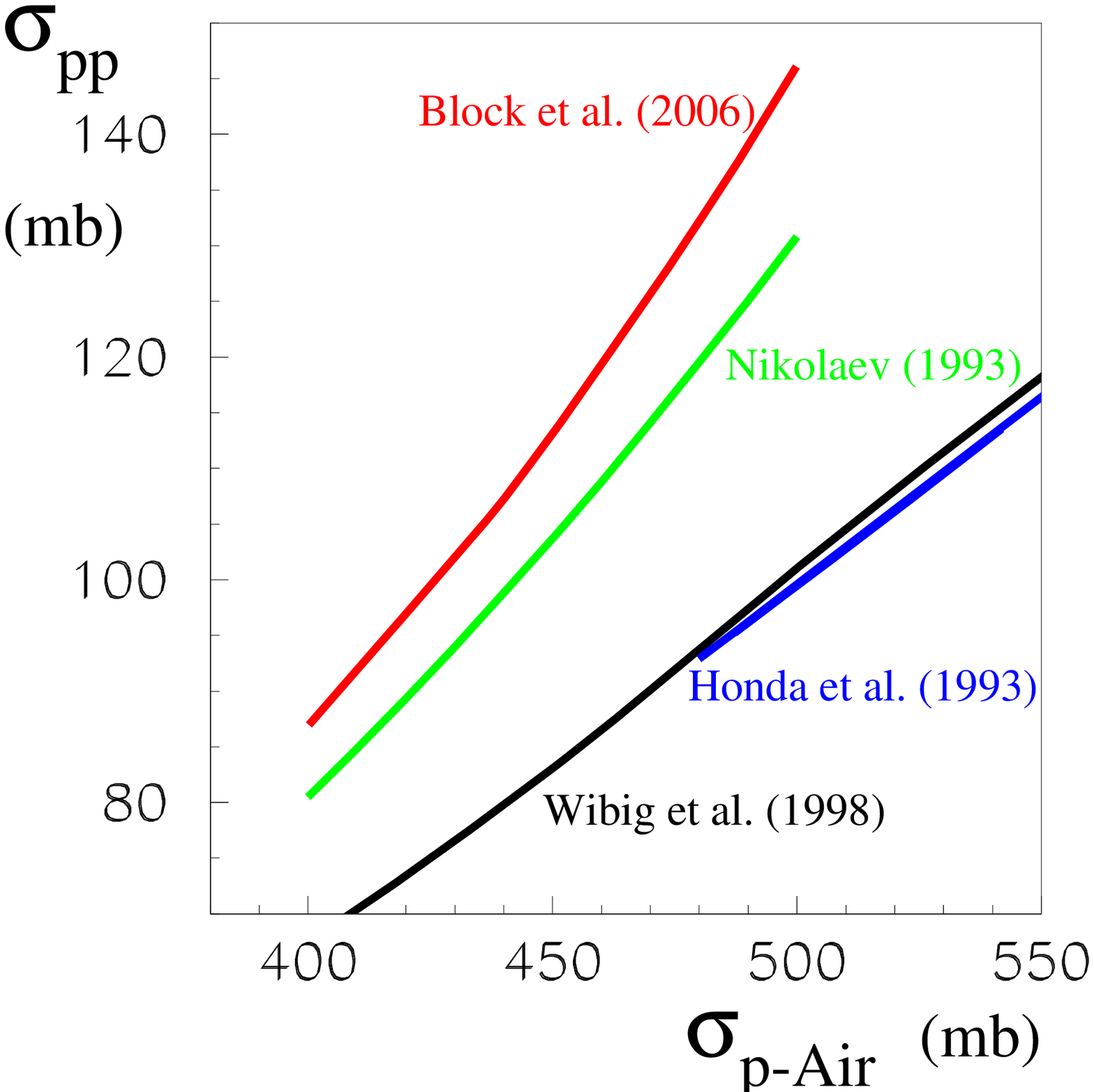}}
\caption{The relation between $pp$ and $P$-Air cross sections calculated
with exact Glauber formula and other dependencies used to convert cosmic ray EAS attenuation data to $pp$ cross section.
\label{pppair}}
\end{figure}
On the other hand, the scattering process
can be treated as the single collision process
with its own nuclear phase shift $\chi_{\rm opt}(b)$
To get 
the consistency with Eq.(\ref{chiskla}) it is required
\begin{eqnarray}
 {\rm e}^{i \chi_{\rm opt}(b)}~=~
\int | \psi(\{{\bf d}\})|^2\:
{\rm e}^{i \sum_{j=1}^A \: \chi _j ({\bf b}\: -\: {\bf d}_j)}
\prod _{j=1}^A d^2 {\bf d}_j
~=
\nonumber \\
~~~~~~~~~~~~~~~~~~~~~~~~~~~~~~
=~\left\langle {\rm e}^{i \chi(b,\: \{{\bf d}\})} \right\rangle~~,
\label{chiopt}
\end{eqnarray}
what defines the relation of the individual projectile-nucleon and 
overall projectile-nucleus oppacities.

To go further with the calculations of $\chi_{\rm opt}$ a commonly used
assumption has to be made. If we assume that the number of scattering centers
($A$) is large and the transparency of the nucleus as a whole remains constant
then
\begin{equation}
\chi_{\rm opt}(b)
~=~i\: \int d^2 {\bf d} \rho_A({\bf d})\:
\left[ 1 - {\rm e}^{i \chi({\bf b} - {\bf d})} \right]~.
\label{exact}
\end{equation}
where $\rho_A$ is the
distribution of scattering center (nucleon) positions
in the nucleus ($\sum \varrho_j$).

\begin{figure}[th]
\centerline{
\includegraphics[width=7.7cm]{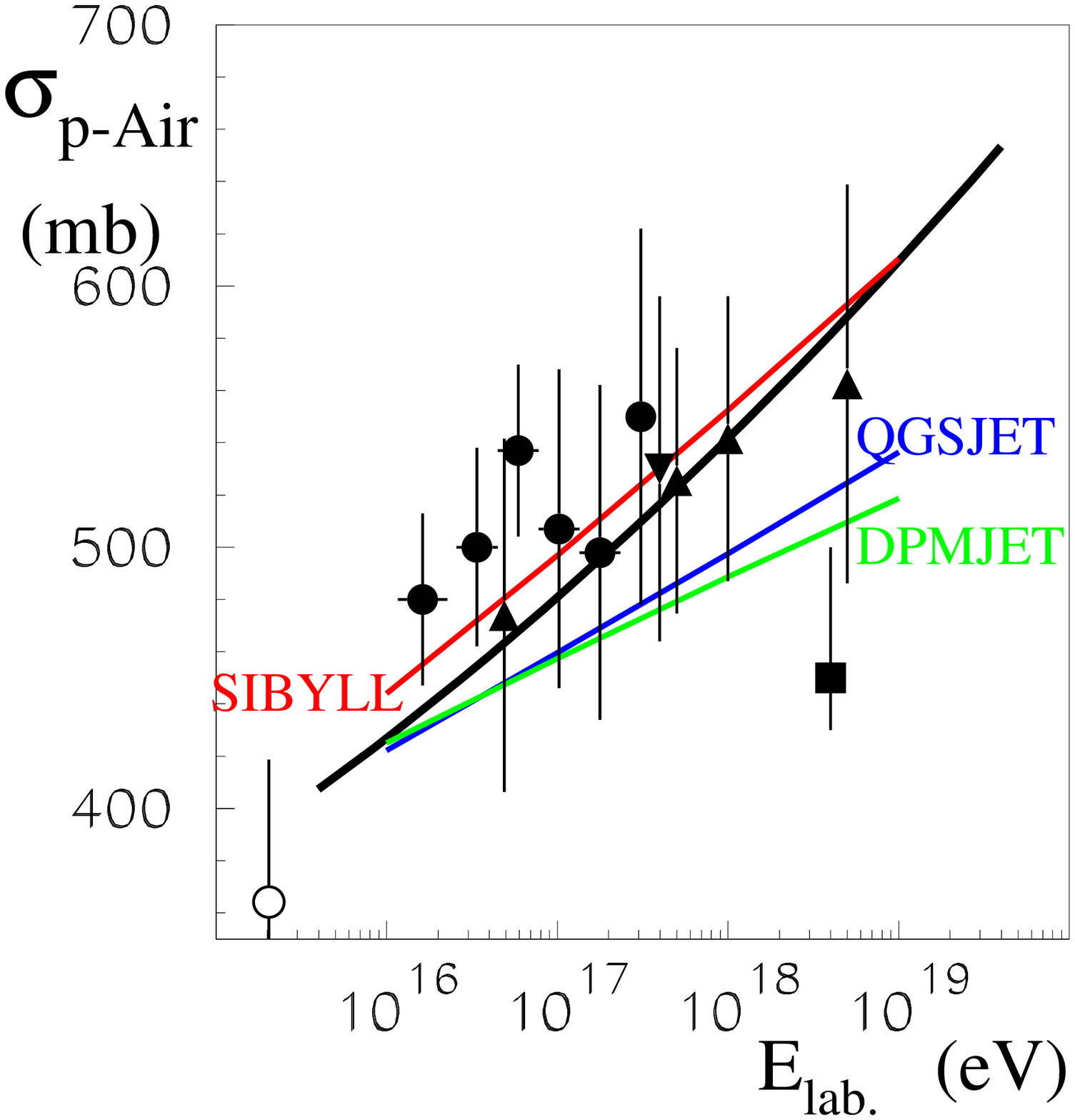}}
\caption{Proton-Air interaction cross section calculated according to
exact Glauber formula and the values used by DPMJET QGSJET and SIBYLL
models of CORSIKA program.
\label{pair}}
\end{figure}

And finally, assuming that
the individual nucleon opacity $| 1-{\rm e}^{i \chi(b)} |$ is a very
sharply peaked compared with $\rho_A$ then with the help of the
optical theorem the simple formula can be found
\begin{eqnarray}
\sigma_{pA}^{\rm inel}~=~
\int d^2 {\bf b}
\left[ 1 - {\rm e}^{- \sigma_{pp}^{\rm tot} \rho_A(b)} \right]
~=~~~~~~~~~~~~~~\nonumber \\
~~~~~~~~~~~~~=~
\int d^2 {\bf b}
\left\{ 1 -
\left[ 1-
\sigma_{pp}^{\rm tot} {\rho_A \over A} \right] ^A \right\}~,
\label{ginel}
\end{eqnarray}
where the last equality holds in the large $A$ limit 
(certainly Eq.(\ref{ginel}) cannot be used for $A=1$)
This result is often but not quite correctly called
``the Glauber approximation''. As it has been shown,
the original Glauber assumption given by Eq.(\ref{chiskla})
is here supported by small nucleon size and a large value of A.

As we saw for very high energies nucleons are quite big objects and
it is expected that at least the last approximation could be questioned.

We have performed respective calculations. The results is shown in the
Fig.\ref{pppair} as a relation between $pp$ and $p$-Air 
interaction cross sections. It is shown there with the other used in the literature.
The importance of this relation is that it allows one to get the $pp$ cross section 
from the cosmic ray data on EAS
attenuation length measured experimentally which is related to $p$-Air
interaction cross section.

The consistency of our cross section description is shown in Fig.~\ref{pair} 
as the $p$-Air cross section calculated with exact Glauber formula Eq.(\ref{chiskla}) 
and $pp$ inelastic cross section parametrization of Eqs.(\ref{lamb}, \ref{inel}). There are 
also shown cross sections adopted by various 
high energy interaction models in CORSIKA.

Concluding the discussion on cross sections, we show in the Table~\ref{lhc} 
some predictions of various authors concerning he $pp$ total cross section
for the LHC energy $\sqrt{s}=14$ TeV.

\begin{table}
\caption{Predicted values of $pp$ cross section at LHC energies.\label{lhc}}
\begin{tabular}{|lcl|l|}
\hline
Author&Year&Ref.&
{$\sigma_{\rm tot}$}\\
\hline
Honda& 1993& \cite{honda} & 110.4 \\
Wibig and Sobczy\'{n}ska &1997 &\cite{wibso}&102.5\\
Cudell {\it et al.} &2002 &\cite{cudell}&111.5$\pm$1.2\\
Velasco {\it et al.}&1999 & \cite{perper99}&104.17$\pm$4.4\\
P\'{e}rez-Peraza {\it et al.} &2005& \cite{perper}&108.27$\ ^{+4.4}_{-3.17}$\\
Block {\it et al.} &1999 &\cite{sigblock}&108$\pm$3.4\\
Block and Halzen & 2006&\cite{blockhalz}&107.3$\pm$1.2\\
Ishida and Igi &2007& \cite{ishida}&109.5$\pm$2.8\\
\hline
\end{tabular}
\end{table}

As it is seen, all the estimated values are very close and, with high degree of 
confidence, it can be concluded that the $pp$ cross section predicted for the LHC 
energies is expected to be equal about 108 mb (within few millibarns 'error box').
The result 102.5 mb is excluded by the most later fits. However, it should be 
underline here again that this result was obtained by 
the exact Glauber formulas while the ~108 mb prediction is influenced by the EAS data
obtained using the transition from $p$-Air to $pp$ made with the help of point nucleon 
approximation or even with multiple scattering approach \cite{gs2}.

\section{'Protons only' and HDPM results on $x_{\rm max}$}

As if is shown in Fig.~6 (left) the pure proton flux around the ankle
is excluded when HDPM with default parameters is used. 
The analytic shower development allows us to test if the correction 
factor of the form of Eq.(\ref{corrfac}) applied to a model parameter
can make the 'pure proton' hypothesis acceptable.

We can try first to change the interaction cross section.
It is possible to get some nice result as it is seen in Fig.~10.

But this results needs cross section as the one shown in the small inserted plot in Fig.~10
what is certainly unacceptable.

Applying the correction given by Eq.(\ref{corrfac}) (with $\delta_{{19}}$ not greater than $\sim 3$) 
to the width of the Gaussian (Fig.~\ref{hdpmpars}),
the average multiplicities (taking care not to exceed the available energy, inelasticity 
coefficient can not be bigger than 1),
or to average $p_\bot$ the 'pure proton' $x_{\rm max}$ can not get close to the measured values. 

\centerline{
\includegraphics[width=7.7cm]{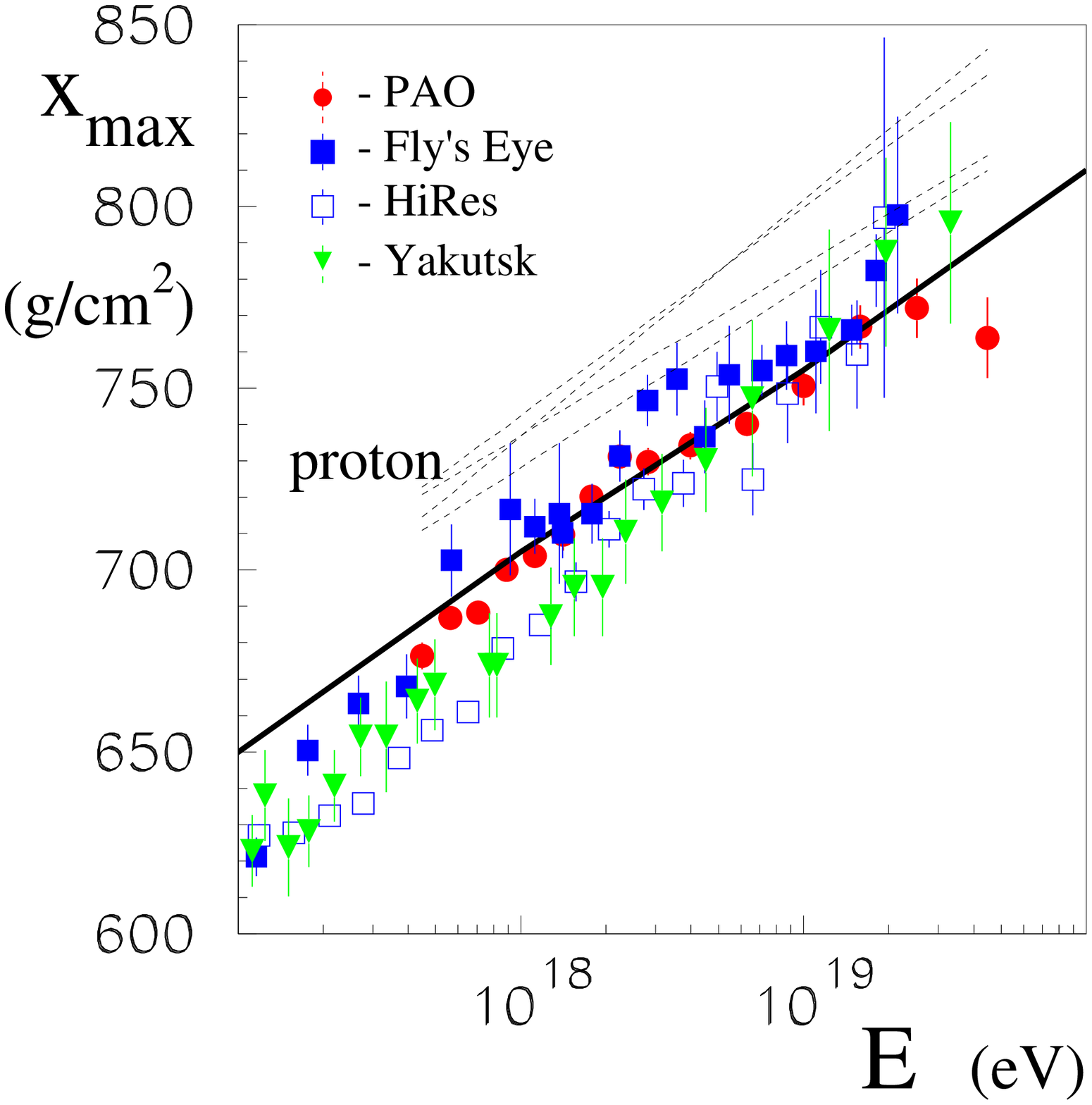}
\hspace{-3.85cm}
\includegraphics[width=3.1cm]{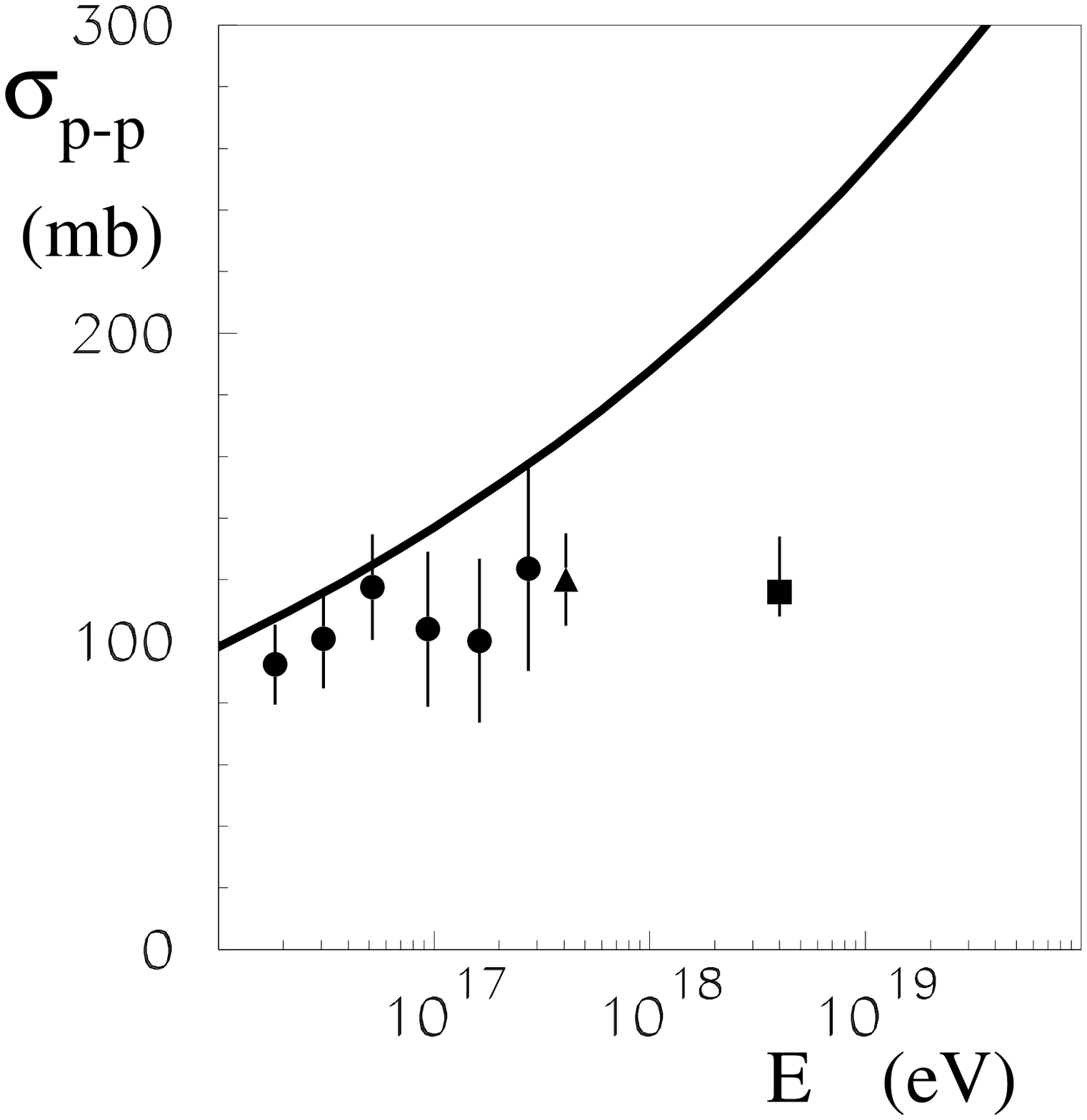}}
{\footnotesize{Fig. 10. HDPM 'best' results for 'pure protons' with the cross section increased. The inserted plot shows a respective increase of 
$\sigma^{pp}_{\rm tot}$ in the very high energy region.}
}
\vspace{.3cm}

Concluding: there is no way to adjust the HDPM parameters in the way that the position of the shower maximum agree with measurements for pure proton composition at and above the ankle.

\vspace{-1cm}
\begin{strip}
\centerline{
\includegraphics[width=7.7cm]{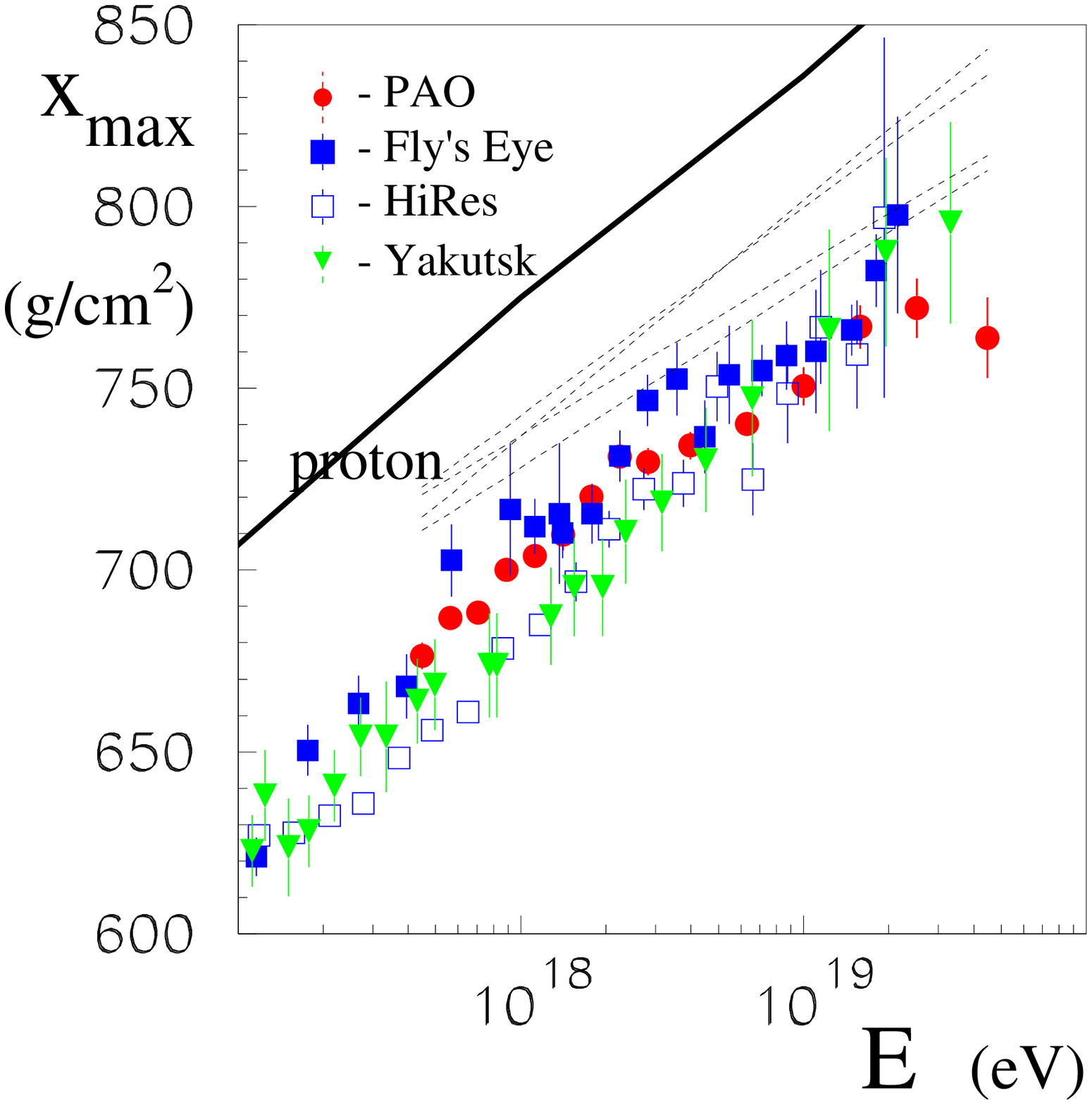}
\hspace{-3.85cm}
\includegraphics[width=3.1cm]{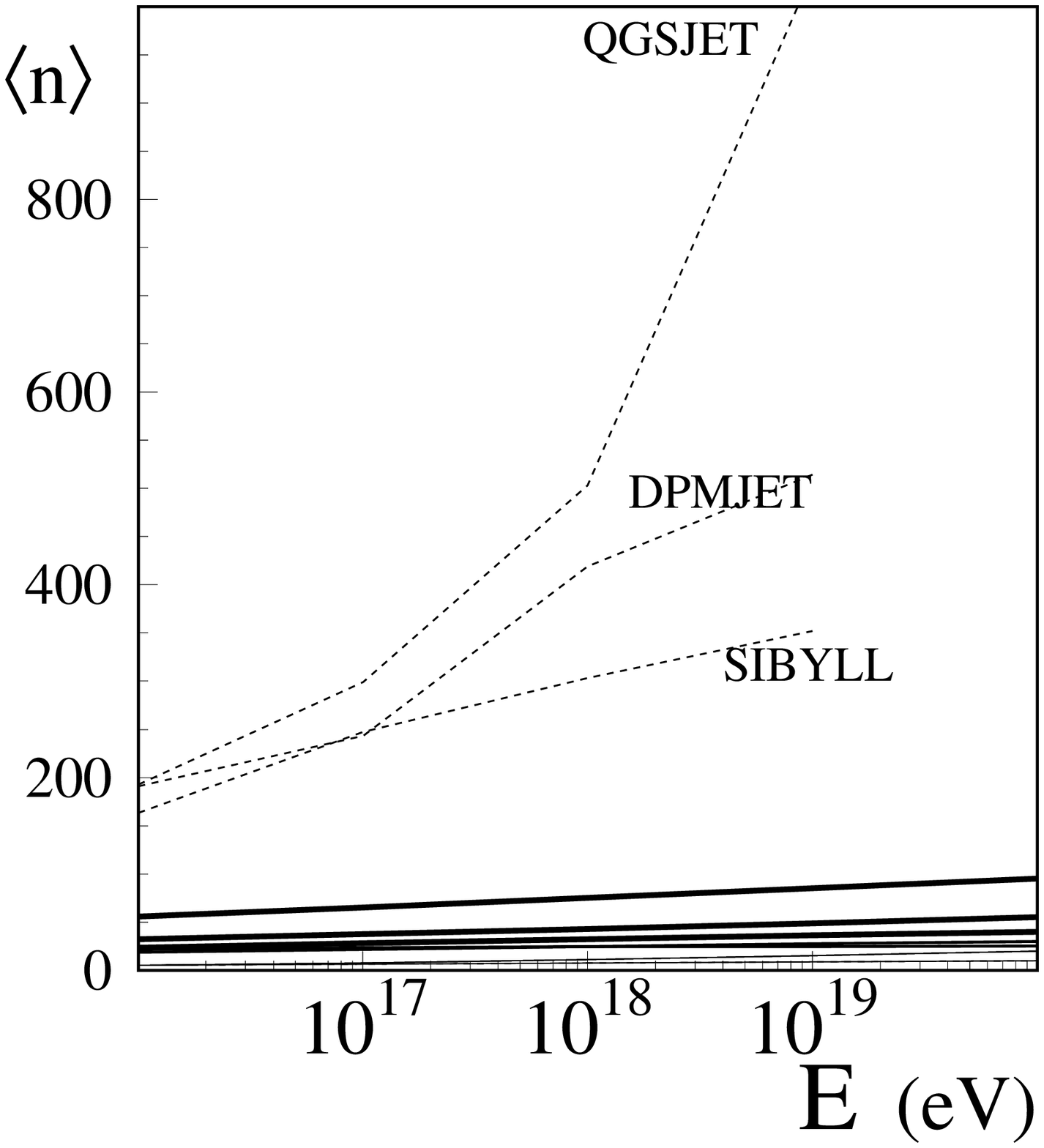}
\hspace{2cm}
\includegraphics[width=7.7cm]{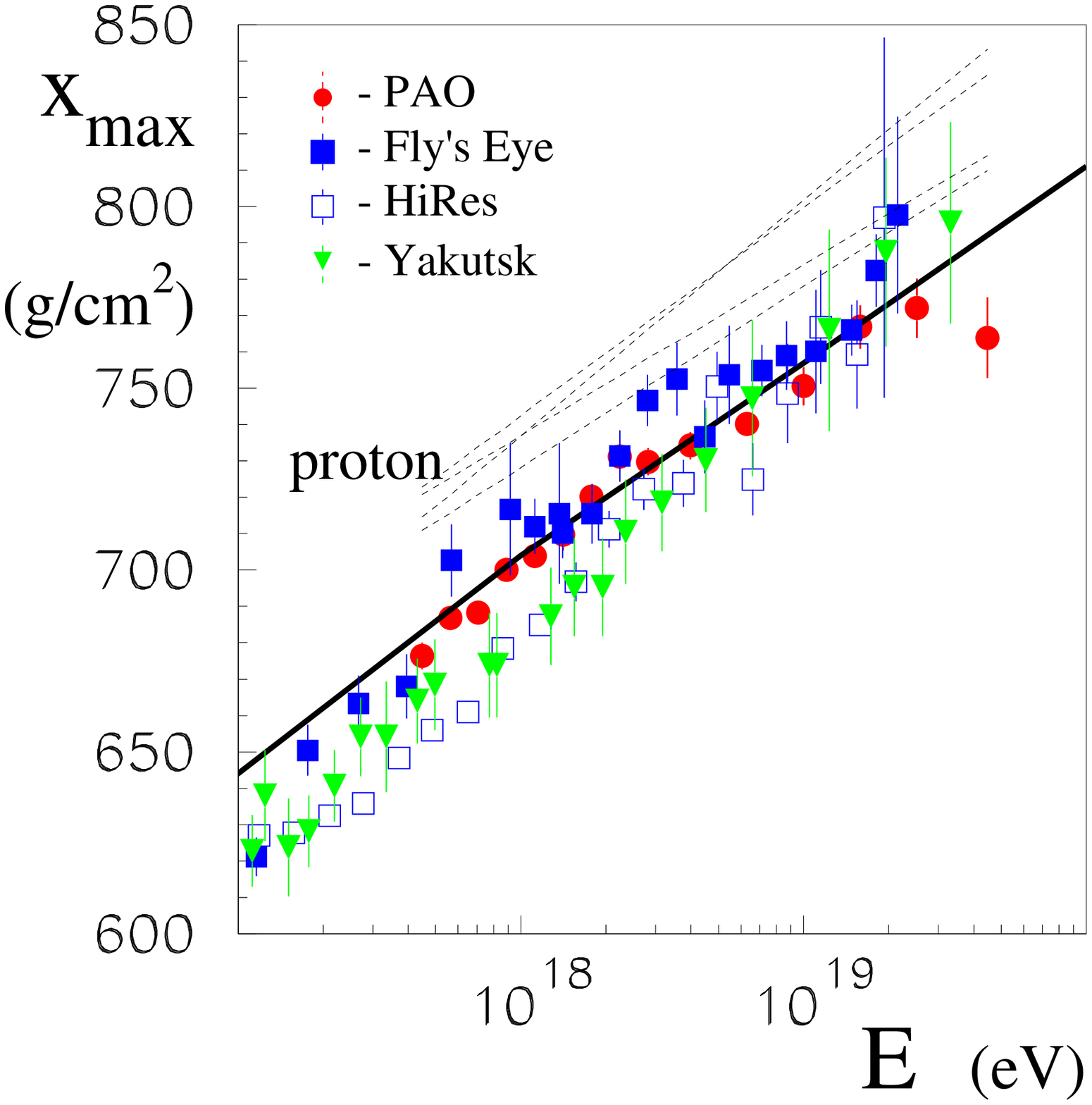}
\hspace{-3.85cm}
\includegraphics[width=3.1cm]{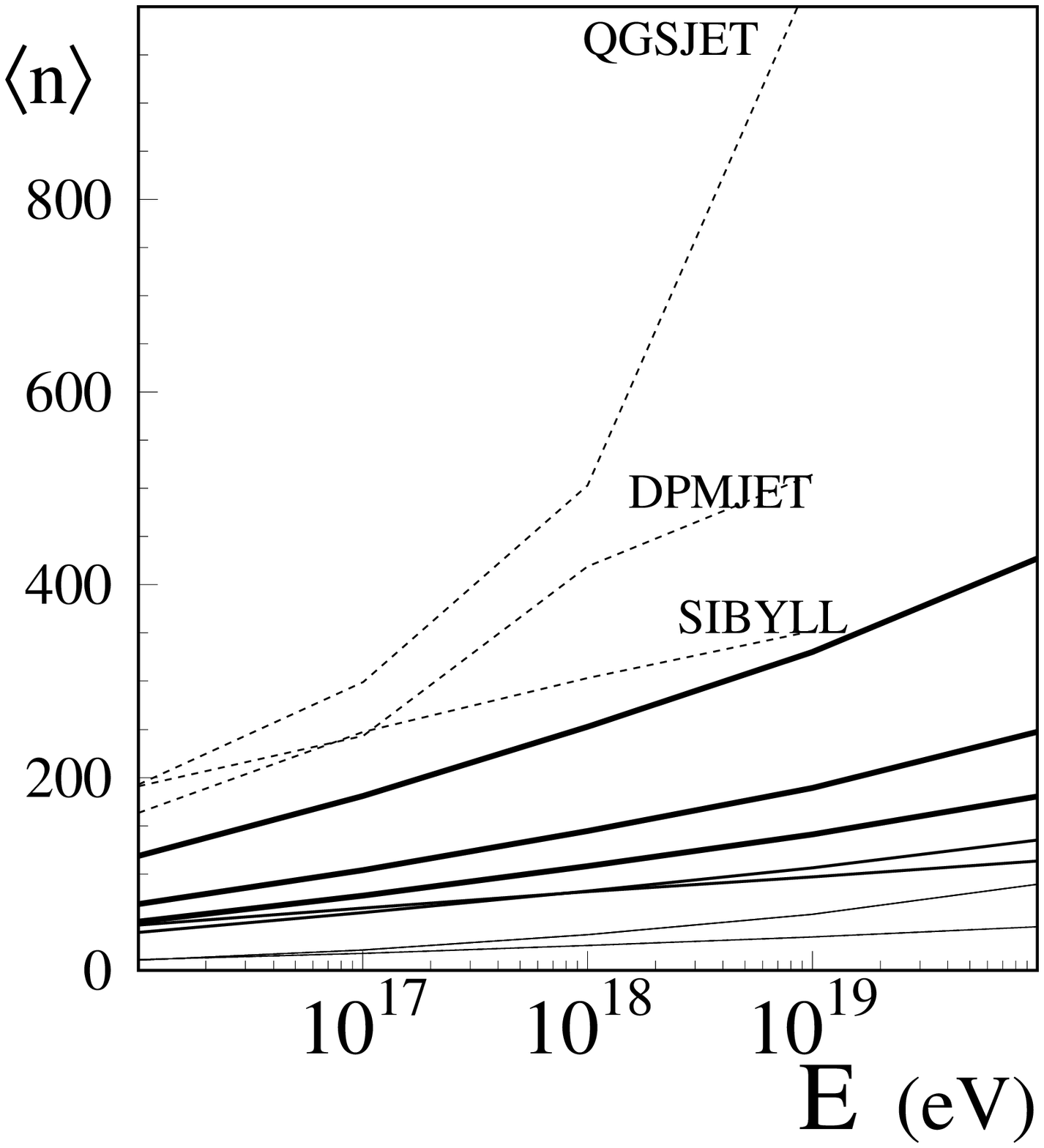}
}
\footnotesize{Fig. 11. Position of the shower maximum as a 
function of primary proton energy calculated with the WW 
model for constant parameter $\alpha$=0 (left) and $\alpha$=0.07 
with slightly increased multiplicity and other parameters 
(right) compared with data.
Small inserted plots shown the average multiplicities respectively.}
\end{strip}

\setcounter{figure}{11}

\section{$x_{\rm max}$ results for 'protons only' with WW model of high energy interaction }

The situation for the Wdowczyk and Wolfendale model is different. 
The proposed strong violation of the Feynman scaling came as a result 
of analysing the variety of cosmic ray data rather than some suggestions 
from simplified theoretical picture of jet fragmentation where the theory 
(QCD-like) has still problems with calculating non-perturbative effects. 
Results shown in Fig.6 give a hope that the WW parametrization could be 
able to describe the position of the shower maximum at very high 
energies with pure proton primary spectrum. The important question 
here is if the eventually adjusted parameters will give 
an acceptable characteristics for the single $pp$ interaction. 

With the help of the fast analytic program the proposed changes to the 
average $p_\bot$, multiplicity and inelasticity has been tested, but the
main interest was put to the value of the WW model parameter 
$\alpha $. Fig.6 shows results for $\alpha$=0.105. 
In Fig.11 results for  
$\alpha=0$ (Feynman scaling) and $\alpha=0.07$.

Values of $x_{\rm max}$ for Feynman scaling are much to deep in the 
atmosphere, and looking into detail of the average multiplicity of 
produced secondaries (shown in Fig. 11 as inserted small plots) one 
can find that they are far too small in comparison with any expectations.

\begin{figure}[th]
\centerline{
\includegraphics[width=7.7cm]{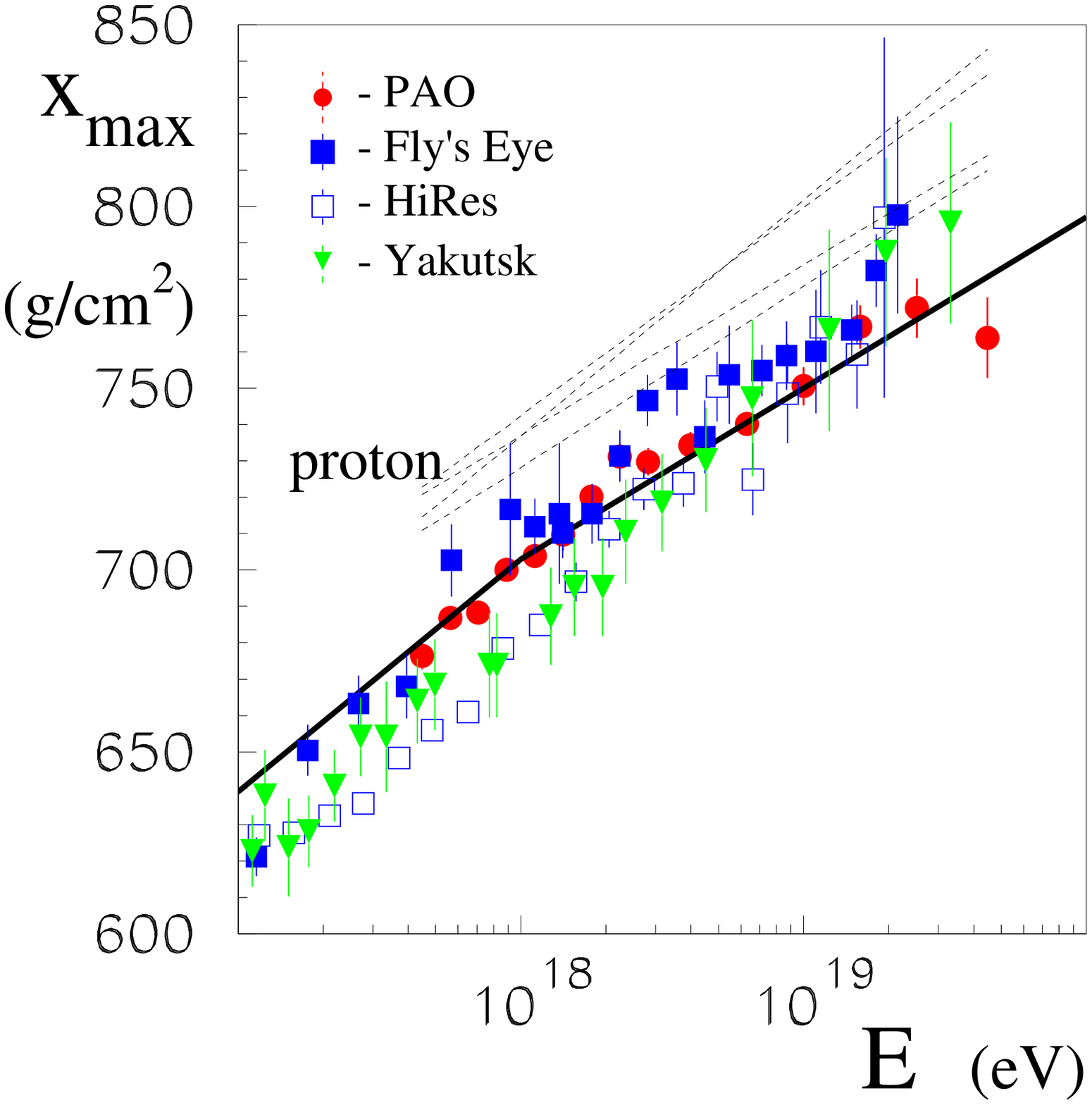}
\hspace{-3.85cm}
\includegraphics[width=3.1cm]{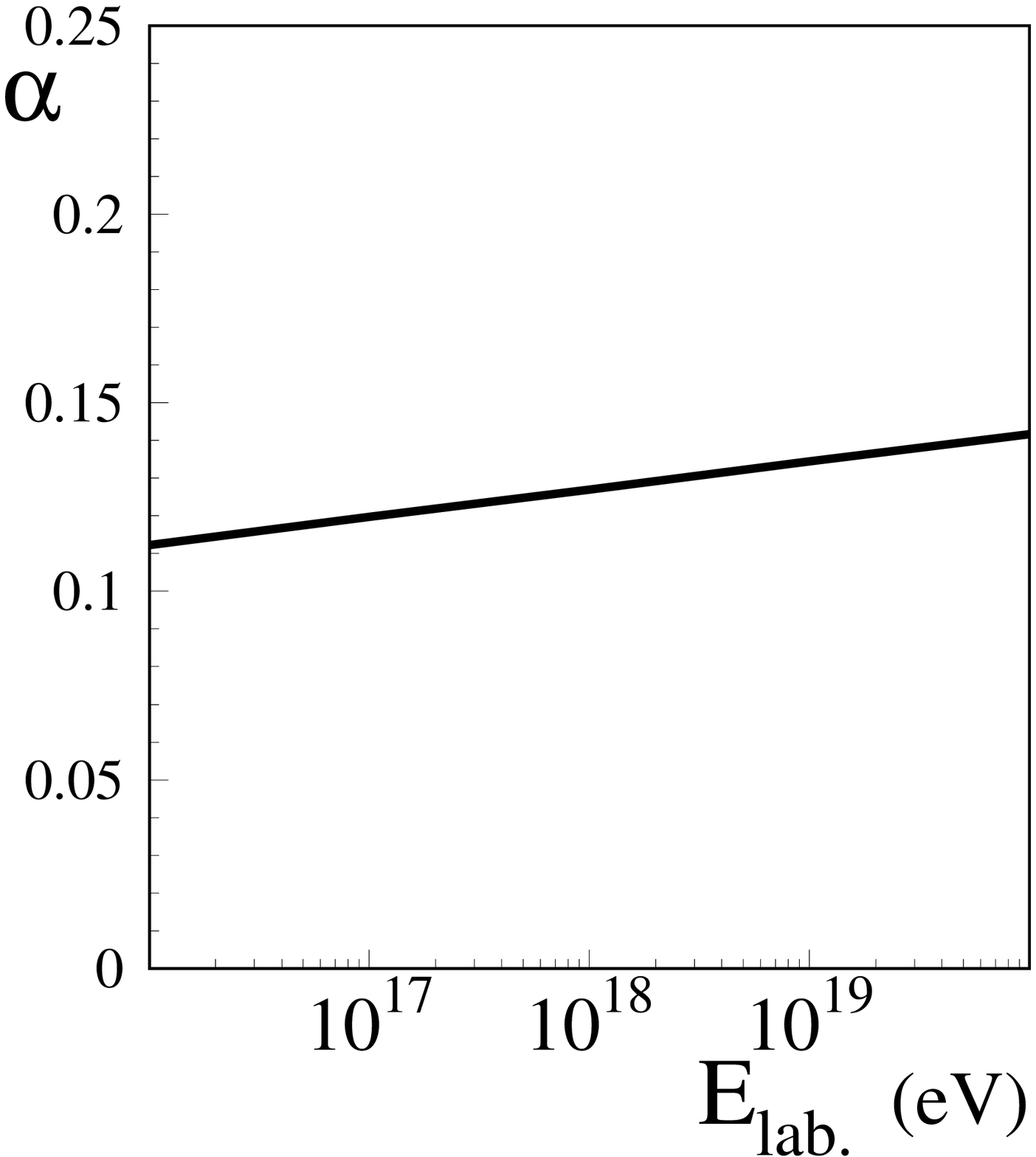}
}
\caption{The 'best' $x_{\rm max}$ result of Wdowczyk and Wolfendale model for 'pure protons'. The inserted plot shows the value of the model parameter
$\alpha$.
\label{xmaxok}}
\end{figure}

The constant $\alpha$=0.07 works better for $x_{\rm max}$ but 
an improvement can be obtained with a changes of the average 
multiplicities according to the form of Eq.(\ref{corrfac}). However,
the multiplicities 
can not be upraised much without a change of secondary particles energy 
distribution because the limited
energy available. The inelasticity calculated by dividing the 
integrated secondaries energies by the incoming energy can reach at most 1 
(the elasticity defined as the fraction energy carried of the most 
energetic particle is of course smaller than 1). Such a limit was 
applied in the analytic integration program, and in this particular case $\rm K_{inel.}$ saturates at 
$\sim 10^{18}$eV

\begin{figure}[th]
\centerline{
\includegraphics[width=5cm]{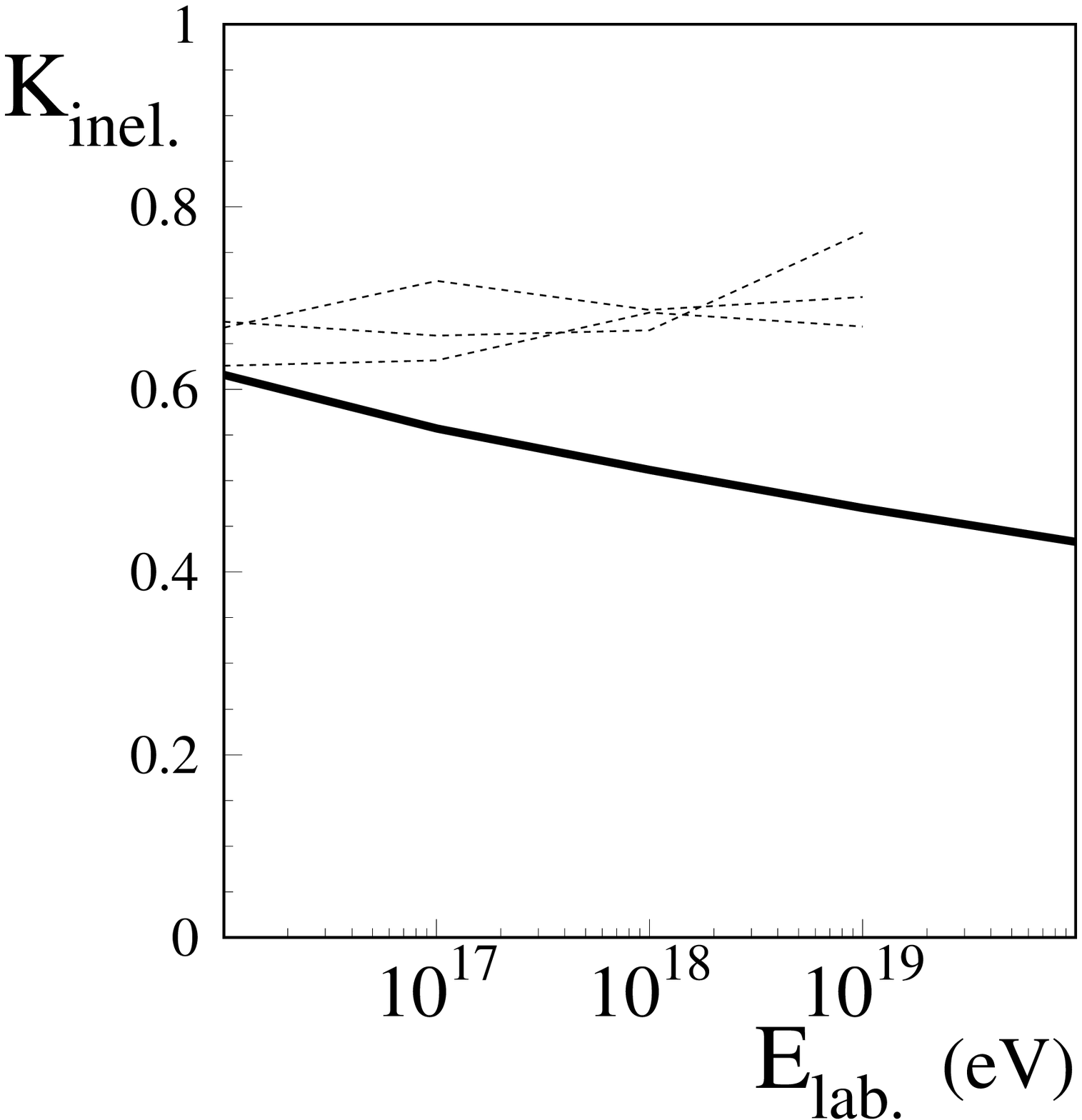}\hspace{-.7cm}
\includegraphics[width=5cm]{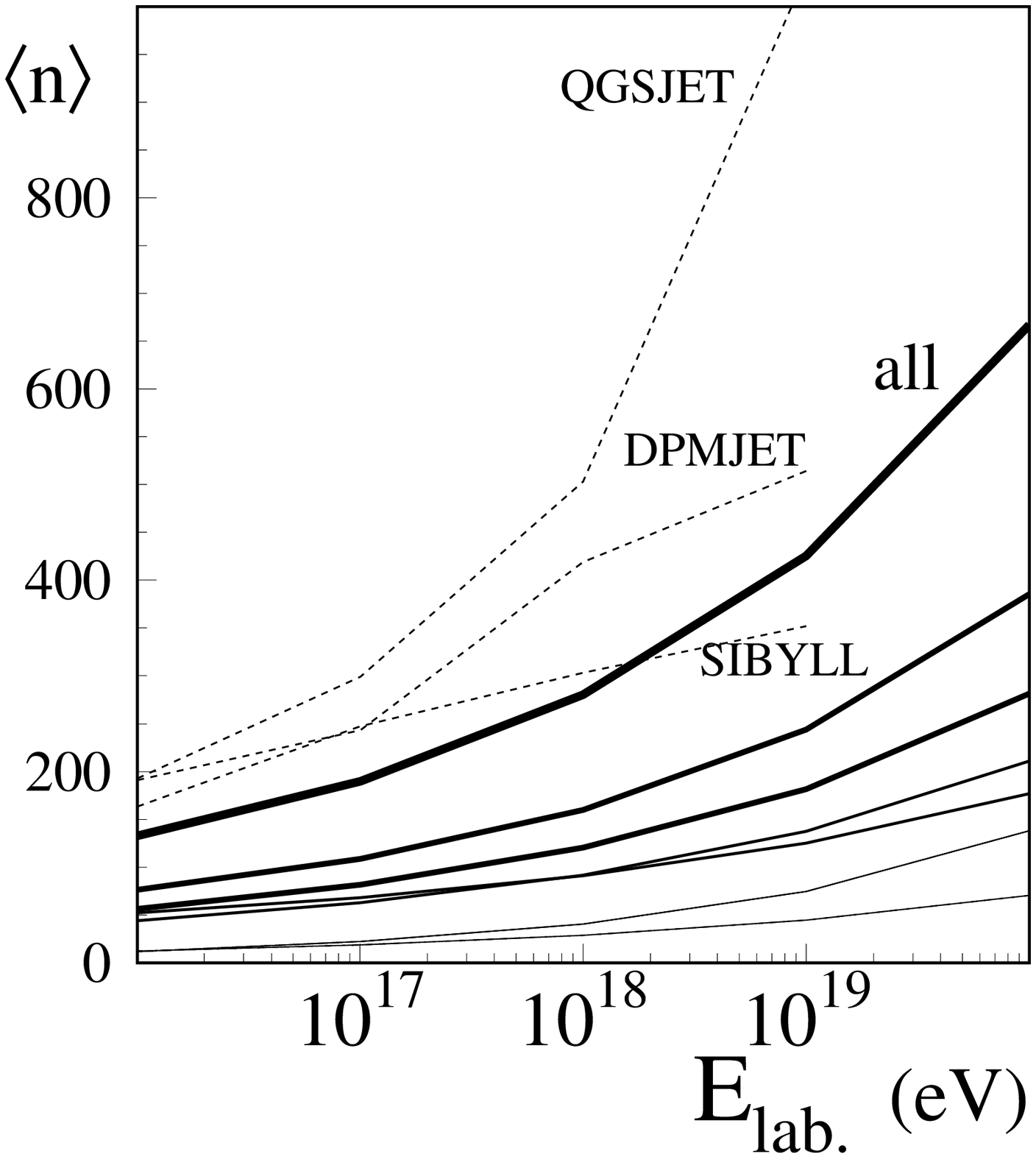}
}
\footnotesize{Fig. 13. Average inelasticity (left)
and multiplicity (right) of the WW model eventually adjusted 
to $x_{\rm max}$ data in Fig.\ref{xmaxok}.}
\end{figure}
\setcounter{figure}{13}

Much better results can be achieved allowing the slow increase of 
the alpha from 0.07 at $10^{15}$eV to $0.134$ at  $10^{19}$eV

Average inelasticity and multiplicities at very high energies of such 'best fit' of $pp$ interaction scaling breaking are shown in Fig. 13. 

\begin{figure}[th]
\centerline{
\includegraphics[width=7.7cm]{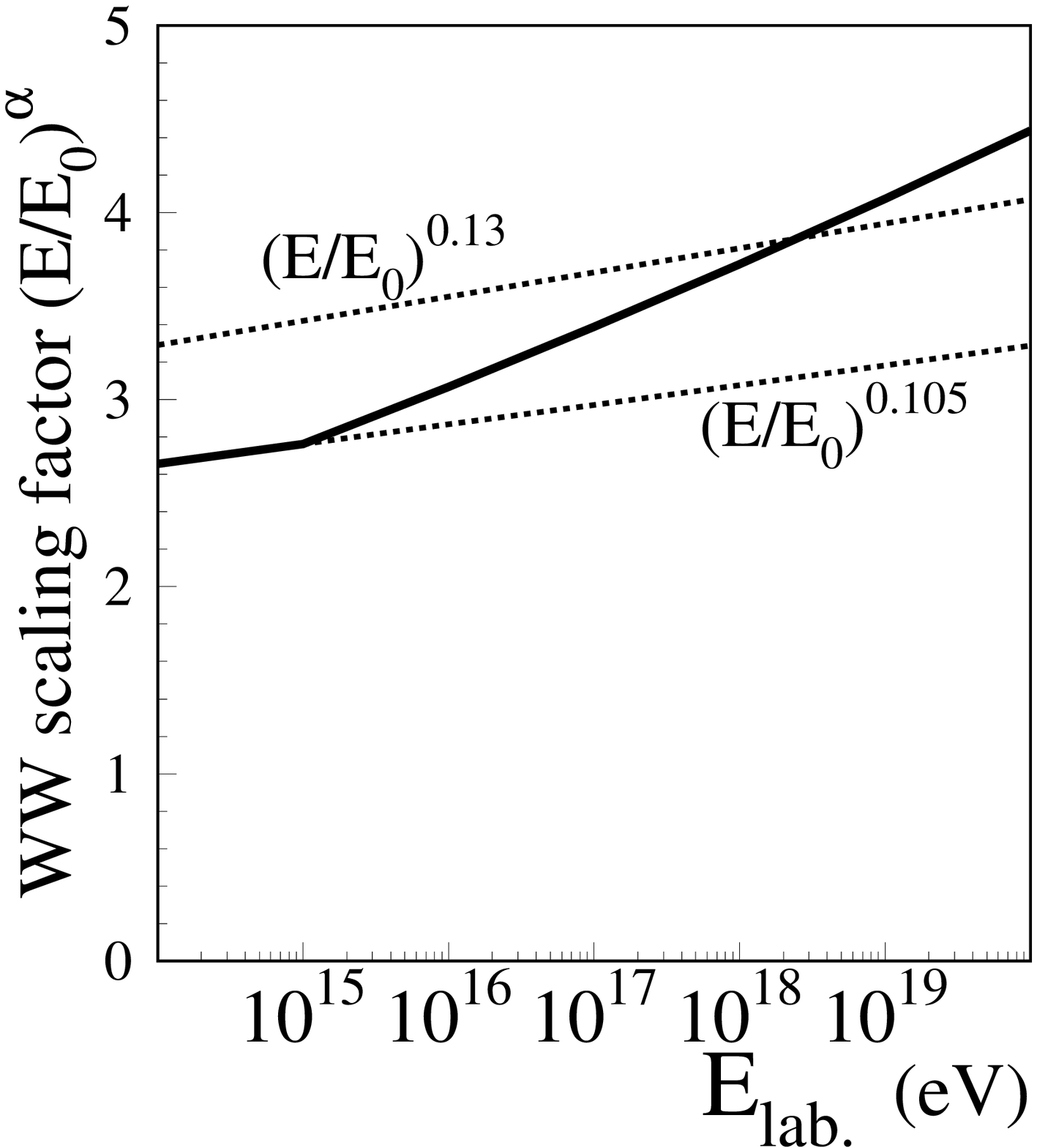}
}
\caption{The WW-scaling factor $\left( {s / s_0}\right)^\alpha$ of the fit shown in Fig.~\ref{xmaxok}.
\label{WWmod}}
\end{figure}

The rise of $\alpha$ can be translated to the behaviour of the effective WW scaling factor $\left( {s / s_0}\right)^\alpha$ modifying the inclusive $x_F$ distributions. This is shown in Fig.\ref{WWmod}.
It eventually goes like the factor of 0.13 found in original Wdowczyk and Wolfendale fits from more than a quarter of century ago.

\section{Conclusions}
We have analyzed the possibility of the pure proton composition of ultra high energy CR 
proposed recently by the PAO and HiRes experiment. This assumption contradicts the conventional wisdom on
the rate of the shower development measured by different big apparatus for some time and 
known as the depth of the shower maximum energy dependence.
The interaction models incorporated in the EAS simulation code CORSIKA all give the composition
enriched by heavies with average logarithm of the mass in about the middle between proton and iron.
It is clear that the pure proton composition needs a change of the interaction model to give the
self-consistent picture of the nature of the CR flux. The way of the changes has been shown. There is 
necessary to violate the Feynman scaling very strongly. We have found that the scaling breaking 
model of Wdowczyk and Wolfendale is well suited to study this. The WW model parameter $\alpha$ which
describes the $x_{\rm max}$ PAO data with the pure proton CR is close to the value found about a 
quarter of century ago by Wdowczyk and Wolfendale in the original papers. The spread of data points and 
reported error bars with possible systematics do not allow us to perform fits much more precise that the 
one presented in this paper. It can be said that the total multiplicities could be slightly below 
expectations, but this is related mostly to the central rapidity region and the WW model is not 
supposed to specify just this region exactly. The average inelasticity could be assumed constant, 
but even slight decrease is possible. 
We have shown also that there is no way to find out the modification of Dual Parton inspired models 
(like HDPM) to be adjusted to the pure proton flux and $x_{\rm max}$ as measured for giant EAS.
If one wants to have protons only than for the interpretation of the shower data, energy calibration etc. 
another interaction model has to be introduced to the CORSIKA repository.


\vspace{.3cm}

\begin{thebibliography}{99}

\bibitem{hage}
R. Hagedorn, {\it Thermodynamics of Strong Interaction}, CERN preprint 1971-12, (1971).

\bibitem{sigmycr}
G.B Yodh et al..Phys. Rev. Lett. {\bf 28} 1005, (1972);

\bibitem{feynman}R.P. Feynman, Phys. Rev. Lett. {\bf 23}, 1415 (1969).
\bibitem{paorec}J. Abraham {\it et al.}, Science {\bf 318} 938, (2007).

\bibitem{hiresxmax}P. Sokolsky and G.B. Thomson, J. Phys. G {\bf 34}, R401 (2007).

\bibitem{corsika}
D. Heck, J. Knapp, J.N. Capdevielle, G. Schatz, T. Thouw, Forschungszentrum Karlsruhe Report, FZKA 6019 (1998). 


\bibitem{jnc}
J.N. Capdevielle, J. Phys. G {\bf 15}, 909 (1989).
\newpage

\bibitem{corsika2}
D. Heck and T. Pierog, {\it Extensive Air Shower Simulation with CORSIKA: A User's Guide
(Version 6.7 from November 19, 2007)}
Institut f\"{u}r Kernphysik Forschungszentrum Karlsruhe.

\bibitem{corsika22}
D. Heck, {\it The influence of hadronic interaction models on simulated air-showers: a phenomenological comparison}, VIHKOS CORSIKA School 2005, Lauterbad.

\bibitem{apel}W. D. Apel {\it et al.}, J. Phys. G. {\bf 34}, 2581 (2007).

\bibitem{paoxmax}M. Unger {\it et al.}, Proc. 30th Internat. Cosmic Ray Conf, Medida, Mexico (2007).

\bibitem{fexmax}D. J. Bird {\it et al.}, ApJ {\bf 424}, 491 (1994).

\bibitem{yakutskxmax}M. N. Dyakonov {\it et al.}, Proc. 23th ICRC, Calgary, Canada, 4, 303 (1993).

\bibitem{ww}J. Wdowczyk and A. W. Wolfendale, Nature {\bf 236}, 29 (1972); \\
J. Wdowczyk and A.W. Wolfendale, J. Phys. A {\bf 6} L48 (1973);\\
J. Olejniczak, J. Wdowczyk and A.W. Wolfendale,  J. Phys. G. {\bf 3} 847(1977);\\
J. Wdowczyk and A.W. Wolfendale, Nuovo Cim {\bf 54}A 433 (1977);\\
J. Wdowczyk and A. W. Wolfendale, Nature {\bf 306}, 347 (1983); \\
J. Wdowczyk and A. W. Wolfendale, J. Phys. G {\bf 10}, 257 (1984).

\bibitem{alner}G.J. Alner {\it et al.} [UA5 Collaboration],
Zeitschrift fur Physik C, 33, 1-6 (1986).

\bibitem{ulrich} R. Ulrich {\it et al.},
{\it On the measurement of the proton-air cross section
using cosmic ray data}, arXiv:0709.1392v1 [astro-ph] (2007).

\bibitem{sigblock}
M. M. Block and F. Halzen, Phys. Rev. D {\bf 63}:114004, (2001);\\
M.M. Block, et al.,  Eur. Phys. J. C {\bf 23} 329, (2002);\\
M. M. Block, et al., Phys. Rev. D {\bf 58}:017503, (1998).

\bibitem{perper}
J. P\`{e}rez-Peraza et al.,NJP  {\bf 7}, 150 (2005).

\bibitem{ishida}
M. Ishida and K. Igi, Eur. Phys. J. C {\bf 52}:357, (2007).

\bibitem{blockhalz}
M. M. Block and F. Halzen, Phys. Rev. D {\bf 72}:036006, (2005).

\bibitem{cudell}
J.R. Cudell {\it et al.} (Compete Coll.), Phys. Rev. Lett. {\bf 89}, 201801 (2002).

\bibitem{wibso}
T. Wibig and D. Sobczynska, J. Phys. G {\bf 24} 2037, (1998). 

\bibitem{blca}
M. M. Block  and R. N. Cahn, Phys. Lett. B {\bf 188}, 143 (1987).

\bibitem{durpi}
L. Durand and H. Pi, Phys. Rev. Lett. {\bf 58}, 303 (1987).

\bibitem{honda}
M. Honda et al. Phys. Rev. Lett.{\bf 70} 525, (1993); 

\bibitem{glauber}
R. Glauber, in Lectures in Theoretical Physics, Eds A. O. Barut and W. E. Brittin, Interscience, New York, (1956);
R. J. Glauber and G. Matthiae, Nucl. Phys. B {\bf 21} 135, (1970).

\bibitem{niko} N. N. Nikolaev, Phys. Rev. {\bf D48} 1904, (1993). 

\bibitem{block2006} M. M. Block, Phys. Rept.{\bf 436} 71, (2006). 

\bibitem{perper99}
J. Velasco {\it et al.}, {\it $\sigma^{pp}_{tot}$ estimations 
at very high energies}, Proc. 20 Internat. Cosmic Ray Conf., 
Utah 1999,  hep-ph/9910484 (1999).

\bibitem{gs2}
J. Engel, T. K. Gaisser, P. Lipari and T. Stanev, Phys. Rev.
{\bf D46}, 5013 (1992).



\bibitem{sigmyaccel}
S.P. Denisov {\it et al.}, Phys. Lett., {\ bf 36} 415, (1971);\\
U. Amaldi {\it et al.}, Phys. Lett. {\bf 44}B 112, (1973).

\bibitem{probAA}
J. Engel, K.T. Gaiser, P. Lipari, and T. Stavev, Phys.Rev.D {\bf 46} 5013, (1992).

\bibitem{mmblock2007}
M.M. Block  {\it Ultra-high Energy Predictions of proton-air Cross Sections from Accelerator Data}, arXiv:0705.3037 (2007).

\bibitem{scalling1}
R. P. Feynman Phys. Rev. Lett. {\bf 23}, 
                             1415 (1969). 

\bibitem{scalling2}  W. Heitler and L. Janossy, 
Proc. Phys. Soc. A{\bf 62}, 662 (1949).




\end{thebibliography}
\end{document}